\documentclass[onecolumn,journal]{IEEEtran}
\usepackage{mathpazo}
\usepackage{times}

\usepackage{amsmath}
\usepackage{amsfonts}
\usepackage{latexsym}
\usepackage{amssymb}
\usepackage{upref}
\usepackage{theorem}
\usepackage{graphicx}
\usepackage{psfrag}
\usepackage{cite}
\usepackage{epsfig}
\usepackage{color}
\usepackage{graphicx}
\usepackage{colortbl}
\usepackage{algpseudocode}
\usepackage{algorithm}

\makeindex

\definecolor{gray}{gray}{0.8}

\theoremstyle{plain}
\theorembodyfont{\normalfont\slshape}

\newtheorem{thm}{Theorem$\!$}

\newtheorem{clm}[thm]{Claim$\!$}

\newtheorem{lem}[thm]{Lemma$\!$}

\newtheorem{prop}[thm]{Proposition$\!$}

\newtheorem{cor}[thm]{Corollary$\!$}

\newtheorem{defn}[thm]{Definition$\!$}

\newtheorem{xmpl}{Example$\!$}

\newtheorem{cnstr}{Construction$\!$}

\newcounter{enumrom}
\renewcommand{\theenumrom}{(\roman{enumrom})}

\makeatletter
\renewcommand{\@endtheorem}{\endtrivlist}
\makeatother

\makeatletter
\renewcommand{\thefigure}{{\@arabic\c@figure}}
\renewcommand{\fnum@figure}{{\bf Figure\,\thefigure}}
\makeatother

\newcommand{\cC}{{\cal C}}

\newcommand{\cM}{{\cal M}}

\DeclareMathOperator{\sgn}{sgn}
\DeclareMathOperator{\spun}{span}

\begin{document}


\title{\textbf{Optimal Rebuilding of Multiple Erasures in MDS Codes}
\vspace*{-0.2ex}}

\author{\IEEEauthorblockN{Zhiying Wang, Itzhak Tamo, and Jehoshua Bruck\\}
}

\maketitle

\begin{abstract}
MDS (maximum distance separable) array codes are widely used in storage systems due to their computationally efficient encoding and decoding procedures. An MDS code with $r$ redundancy nodes can correct any $r$ node erasures by accessing (reading) all the remaining information in the surviving nodes. However, in practice, $e$ erasures is a more likely failure event, for $1 \le e < r$. Hence, a natural question is how much information do we need to access in order to rebuild $e$ storage nodes? We define the \emph{rebuilding ratio} as the fraction of remaining information accessed during the rebuilding of $e$ erasures. In our previous work we constructed MDS codes, called zigzag codes, that achieve the optimal rebuilding ratio of $1/r$ for the rebuilding of any systematic node when $e=1$, however, all the information needs to be accessed for the rebuilding of the parity node erasure.

The (normalized) \emph{repair bandwidth} is defined as the fraction of information transmitted from the remaining nodes during the rebuilding process. For codes that are not necessarily MDS, Dimakis et al. proposed the regenerating codes framework where any $r$ erasures can be corrected by accessing some of the remaining information, and any $e=1$ erasure can be rebuilt from some subsets of surviving nodes with optimal repair bandwidth. 

In this work, we study three questions on rebuilding of codes: (i) We show a fundamental trade-off between the storage size of the node and the repair bandwidth similar to the regenerating codes framework, and show that zigzag codes achieve the optimal rebuilding ratio of $e/r$ for MDS codes, for any $1 \le e \le r$. (ii) We construct systematic codes that achieve optimal rebuilding ratio of $1/r$, for any systematic or parity node erasure. (iii) We present error correction algorithms for zigzag codes, and in particular demonstrate how these codes can be corrected beyond their minimum Hamming distances.
\end{abstract}



\let\thefootnote\relax\footnotetext{
Z. Wang is with Center for Pervasive Communications and Computing, University of California, Irvine, Irvine, CA, USA (email: zhiying@uci.edu).

I. Tamo is with Department of Electrical Engineering - Systems, Tel Aviv University, Tel Aviv, Israel  (email: tamo@post.tau.ac.il).

J. Bruck are with Department of Electrical Engineering, California Institute of Technology,
Pasadena, CA, USA (email: zhiying, bruck@caltech.edu).

The material of the paper was partially presented at the Allerton Conference on Communication, Control, and Computing, 2011 \cite{Wang}.
}

%
\section{Introduction}
%

MDS (maximum distance separable) array codes are a family of erasure-correcting codes used extensively as the basis for RAID storage systems and has been proposed for coding in distributed storage systems. An array code consists of a 2D array where each column can be considered as a storage node. We will use the term column and node interchangeably. An entry in the 2D array is referred to as an \emph{element}. A code with $r$ parity (redundancy) nodes is MDS if and only if it can recover from any $r$ erasures. EVENODD \cite{evenodd} and RDP \cite{RDP} are examples of MDS array codes with two redundancies. In this paper, we only consider  systematic codes, namely, the information is stored exclusively in the first $k$ nodes, and the parities are stored exclusively in the last $r$ nodes.

In order to correct $r$ erasures, it is obvious that one has to access (or read) the information in all the surviving nodes. However, in practice it is more likely to encounter $e$ erasures rather than $r$ erasures, for $1 \le e < r$. So a natural questions is: How much information do we need to access when rebuilding $e$ erasures? Do we have to access all the surviving information? We define the \emph{rebuilding ratio} as the ratio of accessed information to the remaining information in case of $e$ erasures. 

For example, it is easy to check that for the code in Figure \ref{fig1-p}, if any two columns are erased, we can still recover all the information, namely, it is an MDS code. Here all elements are in finite field $\mathbb{F}_3$. Now suppose column $C_1$ is erased, it can be rebuilt by accessing only the elements from rows $0$, $1$ and columns $C_0$, $C_2$, $P_0$, $P_1$ as follows:
\begin{align*}
a_{0,1} &= r_0-a_{0,0}-a_{0,2}=2a_{0,0}+ 2a_{0,2} + r_{0} \\
a_{1,1} &= 2a_{1,0}+2a_{1,2} + r_{1}\\
a_{2,1} &= 2a_{0,0}+2a_{1,2} + z_{0}\\
a_{3,1} &= 2a_{1,0}+a_{0,2} + z_{1}
\end{align*}

Hence, by accessing only $8$ out of $16$ remaining elements, i.e., only half of the remaining information, the erased node can be rebuilt. Similarly, if column $C_0$ or $C_2$ is erased, only half elements need to be accessed. However, if column $P_0$ or $P_1$ is erased, one has to access all elements in column $C_0,C_1,C_2$, a total of $12$ elements, in order to rebuild. Details on this code will be discussed in Section \ref{sec2}.

\begin{figure}
  \centering
  \begin{tabular}{|c|c|c|c||c|c|}
  \hline
  ~& $C_0$ &	  $C_1$ & $C_2$ & $P_0$ & $P_1$ \\
  \hline
  0 & $a_{0,0}$ & $a_{0,1}$ &  $a_{0,2}$ & $r_0=a_{0,0}+a_{0,1}+a_{0,2}$ & $z_0=a_{0,0}+a_{2,1}+a_{1,2}$ \\
  \hline
  1 & $a_{1,0}$ & $a_{1,1}$ &  $a_{1,2}$ & $r_1=a_{1,0}+a_{1,1}+a_{1,2}$ & $z_1=a_{1,0}+a_{3,1}+2a_{0,2}$ \\
  \hline
  2 & $a_{2,0}$ & $a_{2,1}$ &  $a_{2,2}$ & $r_2=a_{2,0}+a_{2,1}+a_{2,2}$ & $z_2=a_{2,0}+2a_{0,1}+2a_{3,2}$ \\
  \hline
  3 & $a_{3,0}$ & $a_{3,1}$ &  $a_{3,2}$ & $r_3=a_{3,0}+a_{3,1}+a_{3,2}$ & $z_3=a_{3,0}+2a_{1,1}+a_{2,2}$ \\
  \hline
  \end{tabular}
  \caption{An MDS array code with three systematic and two parity nodes. All the elements are in finite field $\mathbb{F}_3$. The first parity column $P_0$ is the row sum and the second parity column $P_1$ is generated by the zigzags.} 
  \label{fig1-p}
\end{figure}

The problem of rebuilding information in distributed storage systems has attracted considerable interest in recent years. In the pioneering work \cite{Dimakis2010}, a related problem called {\em repair bandwidth} was proposed. The paradigm there is that one can access the entire information and perform computations within each node, and the question is how much information is {\em transmitted} from some subset of the surviving nodes, called the \emph{helper} nodes, for rebuilding $e=1$ erasure. Moreover, a code is required to reconstruct the entire information from any $r$ erasures, however it is not necessarily MDS, in the sense that the amount of redundant information can be more than the storage size of $r$ nodes. A lower bound on the repair bandwidth was given in \cite{Dimakis2010}. When a single erasure occurs and all the remaining nodes are accessible, the lower bound for the repair bandwidth of an MDS code is $1/r$. It is clear that a lower bound on the repair bandwidth also implies a lower bound on the rebuilding ratio. The MDS codes with optimal repair bandwidth are termed minimum storage regenerating codes in \cite{Dimakis2010}.

A number of codes were designed to achieve the repair bandwidth or rebuilding ratio lower bound for $e=1$ erasure in MDS codes, where all the remaining nodes are helper nodes. When the number of parity nodes is larger than that of the systematic nodes, namely, $r>k$, explicit code constructions were given in \cite{suh2011exact,shah2012interference}. For all cases, \cite{cadambe2013asymptotic} achieved the lower bound asymptotically through interference alignment techniques.
References \cite{Tamo2,viveck} presented explicit constructions of high-rate ($k>r$) MDS array codes  that achieve the lower bound $1/r$ on the rebuilding ratio. Also in \cite{papailiopoulos2013repair} a similar code with $2$ parities was proposed, which has optimal repair bandwidth.
Afterwards,  \cite{Wang2,viveck3,agarwal2015alternate,sasidharan2015high} constructed high-rate codes that achieve optimal repair bandwidth (\cite{Wang2,viveck3}) or optimal rebuilding ratio (\cite{agarwal2015alternate,sasidharan2015high}), where an important consideration was to increase the code dimension, $k$, for a given column length. 

Let us now move from rebuilding a single erasure to multiple erasures. Consider the following example. Suppose that we have an MDS code with $3$ parity nodes. If we have a single erasure, we have codes that achieve the optimal rebuilding  ratio of $1/3$. What happens if we have two erasures? What is the rebuilding ratio in this case? We restrict ourselves to systematic node erasures, whose rebuilding is critical in practice. This is because systematic nodes contain the most frequently requested information from users and requires faster rebuilding than parity nodes.
The first result of this paper is that zigzag codes \cite{Tamo2} can achieve the optimal rebuilding ratio of $2/3$. In general, if we have $r$ parity nodes and $e$ erasures  happen, $1 \le e \le r$,  in our first result of the paper we will prove that the lower bound of repair bandwidth normalized by the size of the remaining array is $e/r$, and so is the rebuilding ratio. In addition, zigzag codes achieve this lower bound for any $e$ systematic node erasures, $1 \le e \le r$.

In our settings, multiple erasures are \emph{simultaneously} rebuilt from information in the remaining nodes.
It should be noted that our model is different from the distributed repair problem, where the recovery of each node is done separately \cite{Dimakis2010}, and multiple erasures simply leads to a smaller number of helper nodes for every single node repair. Also it is different from cooperative recovery schemes \cite{HuCoop}, where a failed node first separately communicate with surviving nodes, and then communicate with each other.

Going back to the single erasure problem, we notice that
the majority of the code constructions for high-rate codes \cite{Tamo2,viveck,papailiopoulos2013repair,Wang2,viveck3,agarwal2015alternate} consider the rebuilding of a systematic node. Reference \cite{sasidharan2015high} addresses the rebuilding of any node erasure, but the code construction is described in a non-systematic fashion. The second result of this paper is a systematic MDS array code construction, such that any node erasure has optimal rebuilding ratio.

We would like to point out here that the constructed code achieves optimal rebuilding ratio at the cost of update complexity. An MDS code with $r$ parities is called \emph{optimal update} if each information element is contained in exactly $r$ parity elements. If we update the value of an information element, we only need to change the value of $r$ parity elements. And this is the minimum number of changes required for an MDS code. For example, in Figure \ref{fig1-p} the information element $a_{0,1}$ is contained in only $r=2$ parity elements: $r_0,z_2$. While the construction in \cite{Tamo2} is optimal update, the constructed code of the second result is not. In fact, each information element is contained in $2r-1$ parity elements.

Finally, in the conventional error model, storage device failures correspond to an erasure or an error of an entire node. Therefore, array codes are usually designed to correct such entire node failures, and the previous work on the rebuilding problem in the literature has focused on recovering from entire node erasures. An exception is partial MDS codes \cite{blaum2013partial} and sector-disk codes \cite{plank2014sector} for RAID systems, which consider the erasure pattern of entire node erasure together with single element erasure. 
The model in \cite{blaum2013partial,plank2014sector} is that an element in the 2D array is a sector in a disk, then by the fault tolerance mechanism in the disks, an element erasure can be detected. Therefore, the code needs to tolerate node and element erasures.
On the other hand, for SSD storage media where symbols can be accessed individually, one can view each element as a single symbol. Then we have the fault type where single element can be erroneous. 
In other words, we may encounter only a few errors in a column as well as entire node erasures.

For an MDS array code with two parities, the minimum Hamming distance is $3$. Therefore, it is not possible to correct a node erasure and a node error at the same time. However, since zigzag code has a very long column length, it is interesting to understand whether it is capable of correcting a node erasure and some element errors.
In the third result of this paper, we show that zigzag code can be a good candidate for correcting node erasures and element errors at the same time.
Moreover, we show how to correct an entire node error for zigzag codes.


The main contributions of this paper are summarized as follows.
\begin{itemize}
  \item
  We establish the lower bound of the rebuilding ratio of $e/r$ for $e$ erasures and $r$ parities. More generally, we show the fundamental tradeoff between the storage cost and repair bandwidth for $e$ erasures similar to the regenerating codes framework.
  \item
  We show an optimal rebuilding algorithm for zigzag codes that achieves the rebuilding ratio of $e/r$ for $e$ systematic erasures.
  \item
  We develop techniques to rebuild zigzag codes (in the generalized sense defined in Section \ref{sec3}), such that a subset of helper nodes access the optimal  $e/r$ fraction of element, and the remaining helper nodes access the entire node information. Based on this technique we present an example code with $r=2$ and close to optimal rebuilding ratio, while the code dimension $k$ is quadratic compared to the previous zigzag code. 
  \item
  A surprising observation is proved: there exists a threshold $e^*$ for an MDS code with $r$ parities, $1 \le e^* \le r$, such that when $e$ erasures occur, $e^* \le e \le r$, the code can be rebuilt optimally, when $e$ satisfies $1 \le e \le e^*$, the code does not have optimal rebuilding ratio.
  \item
  We give an explicit construction of a family of MDS array codes with $r$ parity nodes, that achieves the lower bound $1/r$ for rebuilding {\em any systematic or parity node}. The rebuilding of a single erasure has an efficient implementation as computations within nodes are not required. Moreover, our codes have simple encoding and decoding procedures - when $r=2$ and $r=3$, the codes require finite field sizes of $3$ and $4$, respectively.
  \item
  An algorithm for the zigzag code with $r=2$ is devised  that corrects a single node erasure and an element error. Correcting such fault scenarios demonstrates the potential of zigzag codes to correct beyond its minimum Hamming distance. Besides, an algorithm for correcting a node error is developed for zigzag code with any $r$ parities.
\end{itemize}

The rest of the paper is organized as follows. Section \ref{sec2} introduces the rebuilding problem for MDS array codes and reviews the zigzag code construction \cite{Tamo2}. In Section \ref{sec3} we develop rebuilding algorithm for multiple erasures in general zigzag codes and specifically show that the rebuilding ratio of $e/r$ is a fundamental lower bound and is also achievable by certain zigzag codes. Section \ref{sec5} constructs the codes with optimal rebuilding ratio for both systematic and parity codes. The problem of correcting errors in zigzag codes is shown in Section \ref{sec4}. Finally, the paper is summarized in Section \ref{sec6}.

%
\section{Problem Settings and Zigzag Codes}
\label{sec2}
%
Notations: In the rest of the paper, we are going to use $[i,j]$ to denote $\{i,i+1,\dots,j\}$ and $[i]$ to denote $\{1,2,\dots,i\}$, for integers $i \le j$. 
For a matrix $A$, $A^T$ denotes the transpose of $A$.
For two binary vectors $v=(v_1,\dots,v_m),u=(u_1,\dots,u_m)$, the inner product is $v \cdot u = \sum_{i=1}^{m} v_i u_i \mod 2$. For two permutations $f,g$, denote their composition by $fg$ or $f \circ g$.

In this section we formally define the rebuilding ratio problem and review the zigzag code construction in \cite{Tamo2}, which was shown to have optimal rebuilding for a single systematic node erasure. We then show that the construction can be made an MDS code, in fact, this will be the basis for proving that our newly proposed construction described in Section \ref{sec5} is also an MDS code.

\subsection{Problem Settings}
We first define the framework of a systematic $(n,k)$ MDS array code, which can tolerate  $r$  arbitrary node erasures.
Here $n,k,r=n-k$ are the total length, the dimension, and the number of parities of the code. Let $\cM$ be the size of the information to be stored.
We assume that each systematic node stores $p=\frac{\cM}{k}$ of the information and corresponds to columns $[0,k-1]$. 
Let $A=(a_{i,j})$ be an information array of size $p \times k$. We add $r$ parity columns to this array, such that from any $k$ columns, we can recover the entire information. A column is also called a node, and an entry is called an element. The elements in a column or the rows are indexed $\{0,1,\dots,p-1\}$. 

For a given MDS code with parameters $k,r$, we study the rebuilding of $e$ node erasures in the average case, for $e \le r$. Suppose the possible erasures are from a set of nodes $S$. If $S$ is the set of all $k+r$ nodes, we consider the rebuilding of an arbitrary set of erasures. If $S$ is only the set of $k$ systematic nodes, we consider only the rebuilding of systematic erasures. 
Let $E \subseteq S$ be a subset of $e$ erased nodes. 
The the total number of remaining elements after the erasures of $E$ is $p(n-e)$. Denote by $A(E), T(E)$ be the minimum number of accessed and transmitted elements to rebuild these erasures, respectively.
We define the {\em rebuilding ratio} of a code for $e$ erasures in set $S$ as the fraction of accessed elements for rebuilding:
$$\frac{\sum_{E \subseteq S, |E|=e} A(E)}{\binom{|S|}{e} p(n-e)}. $$
Similarly, generalizing the definition of \cite{Dimakis2010}, we define the (normalized) \emph{repair bandwidth} as the fraction of transmitted information for rebuilding:
$$\frac{\sum_{E \subseteq S, |E|=e} T(E)}{\binom{|S|}{e} p(n-e)}. $$
When the context is clear, we will not mention which set $S$ is in use.
It is easy to see that the repair bandwidth is a lower bound on the rebuilding ratio, because a transmitted symbol can be a function of many accessed symbols. But obviously optimal rebuilding ratio is a more desirable property because it minimizes storage I/O and system cost. Therefore for achievable results of the paper, we focus on optimal rebuilding ratio. We will study repair bandwidth lower bound for multiple erasures in Section \ref{sec3}. 

An MDS code is said to be \emph{optimal update} if each information element is protected by exactly $r$ parity elements. Such codes requires the smallest number of element updates when a single information element needs to be updated.

\subsection{Zigzag Codes}
Next, we review zigzag codes in \cite{Tamo2}.
In \cite{Tamo2}, it was shown that if the code has {optimal update}, then each parity node corresponds to $k$ permutations acting on $[0,p-1]$. More specifically, suppose the permutations are $f_0^l,f_1^l,\dots,f_{k-1}^l$ for the $l$-th parity, $l \in [0,r-1]$. Then the $t$-th element in this parity node is a linear combination of all elements $a_{i,j}$ such that $f_j^l(i)=t$. The set of information elements contained in this linear combination is called a \emph{zigzag set}, denoted by $Z_t^l$. 

Because the ordering of the elements in each node can be arbitrary, we can assume that the first parity node is always a linear combination of each row (corresponding to identity permutations). If we write a permutation in the vector notation, we have
$$f_0^0=f_1^0=\cdots=f_{k-1}^0=(0,1,\dots,p-1).$$
Figure \ref{fig1-p} is an example of such codes. The first parity $P_0$ corresponds to identity permutations, or sum of each row. The second parity $P_1$ corresponds to the permutations
\begin{align*}
  f_0^1 &= (0, 1, 2, 3), \\
  f_1^1&=(2, 3, 0, 1), \\
  f_2^1&=(1, 0, 3, 2).
\end{align*}
To obtain a zigzag set $Z_t^l$, consider for instance, $t=0, l=1$. Since $f_0^1(0)=0, f_1^1(2)=0, f_2^1(1)=0$, the zigzag set $Z_0^1 = \{a_{0,0}, a_{2,1},a_{1,2}\}$, and $z_0$ is a linear combination of these elements.

Consider the rebuilding of systematic nodes in
Figure \ref{fig1-p}. In order to rebuild column $C_1$, we access the
zigzag sets $A=\{Z_0^0,Z_1^0\},B=\{Z_0^1,Z_1^1\}$, corresponding to
parities $\{r_0,r_1\},\{z_0,z_1\}$. Observe that the surviving
elements in $A$ and in $B$ are both $\{a_{0,0},a_{1,0},a_{0,2},a_{1,2}\}$, which are identical and thus have maximal intersection. As a result, only $1/2$ of the elements are accessed. Besides, the coefficients over $\mathbb{F}_3$ in the parity linear combinations guarantee that any two nodes are sufficient to recover all the information, hence the code is MDS.

Now we revisit the set of good permutations such that the accessed zigzag sets have maximum intersection from \cite{Tamo2}. We form permutations based on $r$-ary vectors. Let $e_1,e_2,\dots,e_m$ be the standard vector basis of $\mathbb{Z}_r^m$. Let $e_0$ be the zero vector. We will use $x$ to represent both an integer in $[0,r^m-1]$ and its $r$-ary expansion (the $r$-ary vector of length $m$). It will be clear from the context which meaning is used. All the calculations are done over $\mathbb{Z}_r$. Construction \ref{cnstr5} is a general setup and Theorem \ref{thm_cnstr1} is a special example which achieves optimal rebuilding ratio for systematic nodes. We refer to the former the {\em general zigzag code}, and the latter the {\em optimal zigzag code} throughout the paper. 

\begin{cnstr}[General zigzag code \cite{Tamo2}.]
\label{cnstr5}
Let the information array be $A=(a_{i,j})$ with size $r^m \times k$ for some integers $k,m$.
Let $T=\{v_0,...,v_{k-1}\}\subseteq \mathbb{Z}_r^m$ be a subset of vectors of size $k$, where for each $v=(v_1,...,v_m)\in T,$
\begin{equation}
\gcd(v_1,...,v_m,r)=1,
\label{eq:45678}
\end{equation}
where $\gcd$ is the greatest common divisor. For any $l$, $0\leq l\leq r-1$, and $v\in T$ we define the permutation $f^l_v:[0,r^m-1]\rightarrow [0,r^m-1]$ by $f_v^l(x)=x+lv$, where by abuse of notation we use $x\in[0,r^m-1]$ both to represent the integer and its $r$-ary representation, and all the calculations are done over $\mathbb{Z}_r$.
For simplicity denote the permutation $f^l_{v_j}$ as $f^l_j$ for $v_j\in T$.
For $t \in [0,r^m-1]$, we define the zigzag set in parity node $l$ as 
$Z_t^l=\{a_{i,j}: f^l_j(i)=t\}$.
During rebuilding of systematic node $i$, the elements in rows $X^l_i=\{x\in [0,r^m-1]:x\cdot v_i=r-l\}$ are rebuilt by parity node $l$, $l \in [0,r-1]$.
\end{cnstr}

For example, for $m=2,r=3, x=4, l=2, v=(0,1)$,
$$f_{(0,1)}^2(4)=4+2(0,1)=(1,1)+(0,2)=(1,0)=3.$$
One can check that the permutation $f_{(0,1)}^2$ in a vector notation is $(2,0,1,5,3,4,8,6,7)$.
From \eqref{eq:45678} we get that for any $i \in [0,k-1]$ and $l \in [0,r-1]$,
$|X^l_i|=r^{m-1}$ and it is only $1/r$ of the remaining elements. The following theorem was given in \cite{Tamo2} and shows that in a special case the set $X_i^0$ is sufficient to rebuild node $i$, for any $i \in [0,k-1]$. 

\begin{thm}\label{thm_cnstr1} [Optimal zigzag code \cite{Tamo2}.]
Let the set of $k=m+1$ vectors be $T=\{e_0,\dots,e_m\}$ in Construction \ref{cnstr5}. For the zero vector $e_0$, modify $X_0^l$ to be $\{x\in [0,r^m-1]:x\cdot (1,1,\dots,1)=l\}$. Then the code has optimal rebuilding ratio $1/r$ for rebuilding any systematic node.
\end{thm}

Figure \ref{fig1-p} is an example of Theorem \ref{thm_cnstr1} with $k=3, r=2, m=2$. As mentioned before, only $1/2$ of the information is accessed in order to rebuild $C_1$. The accessed elements are in rows $X_1^0=\{x \in [0,3]: x \cdot e_1=0\}=\{0,1\}$.

Figure \ref{fig5} shows an example with $3$ systematic nodes and $3$ parity nodes constructed by Theorem \ref{thm_cnstr1} with $k=3, r=2, m=2$. The code has an optimal rebuilding ratio of $1/3$. For instance, if column $C_1$ is erased, accessing rows $\{0,1,2\}$ in the remaining nodes will be sufficient for rebuilding.


\begin{figure}
\centering
\begin{tabular}{|c|c|c|}
						\hline
$f_0^1$	&	$f_1^1$	&	$f_2^1$	\\	\hline
0	&	3	&	1	\\	\hline
1	&	4	&	2	\\	\hline
2	&	5	&	0	\\	\hline
3	&	6	&	4	\\	\hline
4	&	7	&	5	\\	\hline
5	&	8	&	3	\\	\hline
6	&	0	&	7	\\	\hline
7	&	1	&	8	\\	\hline
8	&	2	&	6	\\	\hline
\end{tabular}
\hspace{2mm}
\begin{tabular}{|c|c|c|}
						\hline
$f_0^2$	&	$f_1^2$	&	$f_2^2$	\\	\hline
0	&	6	&	2	\\	\hline
1	&	7	&	0	\\	\hline
2	&	8	&	1	\\	\hline
3	&	0	&	5	\\	\hline
4	&	1	&	3	\\	\hline
5	&	2	&	4	\\	\hline
6	&	3	&	8	\\	\hline
7	&	4	&	6	\\	\hline
8	&	5	&	7	\\	\hline
\end{tabular}
\hspace{2mm}
\begin{tabular}{|c|c|c|c|c|c|}
												\hline
$C_0$	&	$C_1$	&	$C_2$	&	$P_0$	&	$P_1$	&	$P_2$	\\	\hline
$a_{0,0}$	&	$a_{0,1}$	&	$a_{0,2}$	&	$a_{0,0}+a_{0,1}+a_{0,2}$	&	$ca_{0,0}+a_{6,1}+a_{2,2}$	&	$c^2a_{0,0}+a_{3,1}+a_{1,2}$	\\	\hline
$a_{1,0}$	&	$a_{1,1}$	&	$a_{1,2}$	&	$a_{1,0}+a_{1,1}+a_{1,2}$	&	$ca_{1,0}+a_{7,1}+ca_{0,2}$	&	$c^2a_{1,0}+a_{4,1}+ca_{2,2}$	\\	\hline
$a_{2,0}$	&	$a_{2,1}$	&	$a_{2,2}$	&	$a_{2,0}+a_{2,1}+a_{2,2}$	&	$ca_{2,0}+a_{8,1}+a_{1,2}$	&	$c^2a_{2,0}+a_{5,1}+ca_{0,2}$	\\	\hline
$a_{3,0}$	&	$a_{3,1}$	&	$a_{3,2}$	&	$a_{3,0}+a_{3,1}+a_{3,2}$	&	$ca_{3,0}+ca_{0,1}+ca_{5,2}$	&	$c^2a_{3,0}+ca_{6,1}+ca_{4,2}$	\\	\hline
$a_{4,0}$	&	$a_{4,1}$	&	$a_{4,2}$	&	$a_{4,0}+a_{4,1}+a_{4,2}$	&	$ca_{4,0}+ca_{1,1}+a_{3,2}$	&	$c^2a_{4,0}+ca_{7,1}+ca_{5,2}$	\\	\hline
$a_{5,0}$	&	$a_{5,1}$	&	$a_{5,2}$	&	$a_{5,0}+a_{5,1}+a_{5,2}$	&	$ca_{5,0}+ca_{2,1}+a_{4,2}$	&	$c^2a_{5,0}+ca_{8,1}+a_{3,2}$	\\	\hline
$a_{6,0}$	&	$a_{6,1}$	&	$a_{6,2}$	&	$a_{6,0}+a_{6,1}+a_{6,2}$	&	$ca_{6,0}+a_{3,1}+a_{8,2}$	&	$c^2a_{6,0}+ca_{0,1}+ca_{7,2}$	\\	\hline
$a_{7,0}$	&	$a_{7,1}$	&	$a_{7,2}$	&	$a_{7,0}+a_{7,1}+a_{7,2}$	&	$ca_{7,0}+a_{4,1}+a_{6,2}$	&	$c^2a_{7,0}+ca_{1,1}+a_{8,2}$	\\	\hline
$a_{8,0}$	&	$a_{8,1}$	&	$a_{8,2}$	&	$a_{8,0}+a_{8,1}+a_{8,2}$	&	$ca_{8,0}+a_{5,1}+ca_{7,2}$	&	$c^2a_{8,0}+ca_{2,1}+ca_{6,2}$	\\	\hline
\end{tabular}
\caption{A $(6,3)$ MDS array code with optimal rebuilding ratio $1/3$. The first parity $P_0$ corresponds to the row sums, and the corresponding identity permutations are omitted. The parities $P_1,P_2$ are generated by the permutations $f_i^1,f_i^2$ respectively, $i=0,1,2$. The elements are from $\mathbb{F}_4$, and $c$ is a primitive element of $\mathbb{F}_4$.}
	\label{fig5}
\end{figure}

\subsection{MDS Property of Zigzag Codes}
Next, we show that by assigning the coefficients in the parities properly, we can make the code MDS.  Our proof techniques are similar to \cite{Tamo2}, which showed the MDS property for zigzag code with $r=2$, and specific coefficient assignments for the optimal zigzag code with $r=2,3$. 

Let $P_j$ be the permutation matrix corresponding to
$f_j=f_j^1$, and denote by $P_j(i,l)$ its $(i,l)$-th entry. Then $P_j(i,l)=1$ if $l+v_j=i$, and $P_j(i,l)=0$
otherwise. Recall that $f_j^1$ is the permutation associated with
parity $1$ and systematic node $j$. 
Next we change the coefficients of the code so that it is MDS, namely, we modify $P_j(i,l)=1$ to some other non-zero value in $\mathbb{F}$. 
When $r=2,3$, assign to the optimal zigzag code for $j \ge 1$
\begin{align}\label{eq925}
  P_j(i,l) = \begin{cases}
  c, & \textrm{if } l \cdot \sum_{t=0}^{j} e_t=0, l+e_j = i,\\
  1, & \textrm{if } l \cdot \sum_{t=0}^{j} e_t\neq 0, l+e_j = i,\\
  0, & \textrm{o.w.}
  \end{cases}
\end{align}
Here $c$ is  a primitive element in $\mathbb{F}_3,\mathbb{F}_4$, for $r=2,3,$ respectively. 
Notice that when $r=2$, this assignment is equivalent to the code construction in \cite{Tamo2}, and when $r=3$ it is identical to \cite{Tamo2}.
For the  general zigzag code, assign
\begin{align}\label{eq926}
  P_j(i,l) = \begin{cases}
  \lambda_j, & \textrm{if } l+v_j = i,\\
  0, & \textrm{o.w.}
  \end{cases}
\end{align}
One can see that in all cases, $P_j^s$ is a generalized permutation matrix corresponding to $f_j^s$.


Let the generator matrix of the code be
\begin{equation}\label{eqG}
G=\left(
\begin{array}{ccc}
I & & \\
& \ddots & \\
& & I \\
I & \cdots & I \\
P_0^1 & \cdots & P_{k-1}^1 \\
\vdots & & \vdots \\
P_0^{r-1} & \cdots & P_{k-1}^{r-1}
\end{array}
\right)_{(k+r)\times k}.
\end{equation}
Here each sub-matrix is of size $r^m \times r^m$. Let the information be arranged as a row vector $a=(a_0,a_1,\dots,a_{k-1})$ of length $r^m k$ and the codeword be arranged as $b=(b_0,b_1,\dots,b_{n-1})$ of length $ r^m n$, where each $a_i$ or $b_i$ corresponding to node $i$ is a vector of length $r^m$. Then we compute the codeword as$$b=Ga.$$ We say that matrix $P_j^t$ is the \emph{encoding matrix} for systematic node $j$ and parity node $t$.

{\bf Remark:} In general, suppose $A_{j,t}$ is the encoding matrix for systematic node $j$ and parity node $t$. We can view every parity element as an equation. Then the $(i,l)$-th entry in the matrix corresponds to the coefficient for the $i$-th element (or equation) in parity $t$, and the $l$-th element (or variable) in systematic node $j$.

For example, the coefficients in Figure \ref{fig1-p} is assigned according to \eqref{eq925}, with
$$P_1 = \left(\begin{array}{cccc}
0&0&1&0 \\
0&0&0&1 \\
2&0&0&0 \\
0&2&0&0
\end{array}\right), P_2=\left(\begin{array}{cccc}
0&1&0&0 \\
2&0&0&0 \\
0&0&0&2 \\
0&0&1&0
\end{array}\right).
$$

The following theorem shows that using the above assignments the code can be MDS.

\begin{thm}
\label{thm1}
(i) Construction \ref{cnstr5} can be made an MDS code for a large enough finite field. \\
(ii) When $r=2,3$, field of size $3$ and $4$ is sufficient to make the optimal zigzag code in Theorem \ref{thm_cnstr1} MDS.
\end{thm}
\begin{IEEEproof}
Part (i): An MDS code means that it can recover any $r$ erasures. Suppose $t$ systematic nodes and $r-t$ parity nodes are erased, $1 \le t \le r$. Thus suppose we delete from $G$ in \eqref{eqG} the systematic rows $\{j_1,j_2,\dots,j_t\}$ and the remaining parity nodes are $\{i_1,i_2,\dots,i_t\}$. Then the following $t \times t$ block matrix should be invertible:
\begin{equation}
\label{eq3}
  G'=\left(
  \begin{array}{ccc}
    P_{j_1}^{i_1} & \cdots & P_{j_t}^{i_1} \\
    \vdots & & \vdots \\
    P_{j_1}^{i_t} & \cdots & P_{j_t}^{i_t}
  \end{array}
  \right).
\end{equation}
Its determinant $\det(G')$ is a polynomial with indeterminates $\lambda_{j_1},\dots,\lambda_{j_t}$. All terms in the polynomial have highest degree $r^m (i_1+\dots+i_t)$. One term with highest degree is $\prod_{s=1}^{t}\lambda_{j_s}^{i_s r^m}$ with non-zero coefficient $1$ or $-1$. So $\det(G')$ is a non-zero polynomial. Up to now we only showed one possible case of erasures. For any $r$ erasures, we can find the corresponding $\det(G')$ as a non-zero polynomial. The product of all these polynomials is again a non-zero polynomial. Hence by \cite{Alon} for a large enough field there exist assignments of $\{\lambda_j\}$ such that the evaluation of the polynomial is not $0$. Then for any case of $r$ erasures, the corresponding matrix $G'$ is invertible, and the code is MDS.

Part (ii) was shown in \cite{Tamo2} for the assignment in \eqref{eq925}.
\end{IEEEproof}

{\bf Remark:} If the non-zero elements in $P_j$ are assigned to be different indeterminates, say, $P_j(i,l)=\lambda_{j,i,l}$ if $l+v_j=i$, and $P_j(i,l)=0$ otherwise. Then the above proof still works and in particular, every determinant $\det(G')$ is a non-zero polynomial.


\section{Rebuilding Multiple Erasures}
\label{sec3}

In this section, we discuss the rebuilding of $e$ erasures, $1 \le e \le r$. In order to simplify some of the results we will assume that $r$ is a prime and the calculations are done over $\mathbb{F}_r$. Note that all the result can be generalized with minor changes for an arbitrary integer $r$ and the ring $\mathbb{Z}_r$.

We will first prove the lower bound for rebuilding ratio and repair bandwidth. Then we show a construction achieving the lower bound for systematic nodes. At last we show some

\subsection{Lower Bounds}

We first show that the rebuilding ratio for zigzag codes is at least $e/r$. Then we show that this lower bound is a special case for a more general class of codes. 

\begin{thm} \label{thm0419}
Consider a zigzag code with $r$ parity nodes. In an erasure of $1\leq e \leq r$ systematic nodes, the rebuilding ratio is at least $\frac{e}{r}$.
\end{thm}

\begin{IEEEproof}
Let the information array be of size $p \times k$. From one zigzag set corresponding to some parity, we get one linear equation corresponding to the erased elements. 
Thus in order to recover the $e\cdot p$ elements in the systematic nodes we need to use at least $e p$ zigzag sets from the $r$ parities.
By the pigeonhole principle there is at least one parity node, such that at least $ep/r$ of its zigzag sets are used. Hence each remaining systematic node has to access its elements that are contained in these zigzag sets. Therefore each systematic node accesses at least $ep/r$ of its information elements, which is a portion of $\frac{e}{r}.$

Since we use at least $ep$ zigzag sets, we use at least $ep$ elements in the $r$ parity nodes, which is again a portion of $\frac{e}{r}$. Hence the overall rebuilding ratio is at least $\frac{e}{r}$.
\end{IEEEproof}

Next, we study the repair bandwidth, namely, the amount of information needed to transmit in order to rebuild $e$ nodes for a general code (not necessary MDS, systematic, or optimal update).
More precisely, we study \emph{exact-repair codes} that satisfy the following two properties: (i) Reconstruction: any $k$ nodes can rebuild the total information. (ii) Exact repair: if $e$ nodes are erased, they can be recovered exactly by transmitting information from the remaining nodes. 
The total amount of information is denoted by $\cM$, and assume the $n$ nodes are indexed by $\{1,2,\dots,n\}$. For $e$ erasures, $1\leq e \leq r$, denote by $\alpha, d_e,\beta_e$ the amount of information stored in each node, the number of helper nodes to repair the erased nodes, and the amount of information transmitted by each of the helper nodes, respectively. Assume that $d_e \ge e$. We want to derive the trade-off between these parameters for an exact-repair code.

The above definitions are extensions of the single-erasure exact-repair codes in \cite{KumarProof}.  The following results give a lower bound of the repair bandwidth for $e$ erasures, and the proof is a generalization of \cite{KumarProof}.

To prove the main lower bound result, let us bound the conditional entropy of two nodes. We first define some random variables corresponding to the stored and transmitted information.
For subsets $A,B \subseteq [n]$, let $W_A$ be the information stored in nodes $A$, and $S_A^B$ be the information transmitted from nodes $A$ to nodes $B$ in the rebuilding.

\begin{lem}
Let $B\subseteq [n]$ be a subset of nodes of size $e$, then for an arbitrary set of nodes $A$, $|A|\leq d_e$, such that $B\cap A=\emptyset$,
$$H(W_B|W_A)\leq \min\{|B|\alpha,(d_e-|A|)\beta_e\}.$$
\end{lem}

\begin{IEEEproof}
If nodes $B$ are erased, consider the case of having nodes $A$ and nodes $C$ as helper nodes, $|C|=d_e-|A|.$ Then the exact repair condition requires
\begin{align*}
0&=H(W_B|S_A^B,S_C^B)\\
&=H(W_B|S_A^B)-I(W_B,S_C^B|S_A^B)\\
&\geq H(W_B|S_A^B)-H(S_C^B)\\
&\geq H(W_B|S_A^B) -(d-|A|)\beta_e\\
&\geq H(W_B|W_A) -(d-|A|)\beta_e .
\end{align*}
Moreover, it is clear that $H(W_B|W_A)\leq H(W_B)\leq |B|\alpha$ and the result follows.
\end{IEEEproof}

\begin{thm} \label{th0503}
Any exact-repair code with file size  $\cM$ must satisfy that for any $1\leq e \leq r$, 
$$\cM\leq s \alpha+ \sum_{i=0}^{\lfloor \frac{k}{e}\rfloor-1} \min\{e \alpha,(d_e-ie-s)\beta_e\}$$
where $s=k \mod e, 0 \le s < e$. Moreover for an MDS code, if $e \le k$,
$$\beta_e \geq \frac{e\cM}{k(d_e-k+e)}.$$
\end{thm}

\begin{IEEEproof}
The  file can be reconstructed from any set of $k$ nodes, hence
\begin{align*}
\cM &=H(W_{[k]})\\
&=H(W_{[s]})+\sum_{i=0}^{\lfloor\frac{k}{e}\rfloor-1}H(W_{[ie+s+1,(i+1)e+s]}|W_{[ie+s]})\\
&\leq s \alpha+ \sum_{i=0}^{\lfloor \frac{k}{e}\rfloor-1} \min\{e \alpha,(d_e-ie-s)\beta_e\}.
\end{align*}
In an MDS code $\alpha=\frac{\cM}{k}$, hence in order to satisfy the inequality any summand of the form $\min\{e\alpha,(d_e-ie-s)\beta_e\}$ must be at least $\frac{\cM-s\alpha}{\lfloor \frac{k}{e} \rfloor} =e\frac{\cM}{k}$, which occurs if and only if
$(d_e-(\lfloor \frac{k}{e} \rfloor-1 )e-s)\beta_e\geq \frac{e\cM}{k}$. Hence we get $$\beta_e \geq \frac{e\cM}{k(d_e-k+e)}.$$
And the proof is completed.
\end{IEEEproof}

Therefore, the lower bound of the normalized repair bandwidth for an MDS code (not necessarily optimal update) with $d_e=n-e$ is 
$$ \frac{e\cM}{k(d_e-k+e)} \cdot \frac{k}{\cM}=\frac{e}{r},$$ 
which is the same as the lower bound of the rebuilding ratio in Theorem \ref{thm0419}.

\subsection{Rebuilding Algorithms}
Theorems \ref{thm0419} and \ref{th0503} provide a lower bound of $e/r$ on the rebuilding ratio in case of $e$ erasures, $1 \le e \le r$.
In this subsection we discuss how to rebuild an MDS array code with optimal update. 
The first main result is Theorem \ref{thm0120}, which states that the optimal zigzag code in Theorem \ref{thm_cnstr1} achieves the optimal rebuilding ratio of $e/r$. 
In order to prove this optimality result, we first prove necessary and sufficient conditions for optimal rebuilding in Lemmas \ref{lem0120} and  \ref{thm03311}, based on which we prove sufficient conditions of optimal rebuilding in Lemmas \ref{thm0413}, \ref{thm0414}.
Correspondingly, Algorithm \ref{alg0120} specifies the rebuilding strategy that satisfies the sufficient conditions.
The second main result is that the derived conditions in Lemmas \ref{lem0120}\ref{thm03311} \ref{thm0413} and \ref{thm0414} holds for non-optimal codes as well if we consider the rebuilding ratio incurred at every individual helper node;
and Algorithm \ref{alg0120} work for general zigzag codes as well, and provides a potential trade-off between the code dimension $k$ and the rebuilding ratio, for given $r,p$. 

We start with an example that achieves the optimal rebuilding ratio, which is in fact an instance of the optimal zigzag code in Theorem \ref{thm0120}. 

\begin{xmpl} \label{xmpl0503}
Consider the code in Figure \ref{fig5} with $r=3$. When $e=2$ and columns $C_0,C_1$ are erased, we can access rows $\{0,1,3,4,6,7\}$ in column $C_2,P_0$, rows $\{1,2,4,5,7,8\}$ in column $P_1$, and rows $\{2,0,5,3,8,6\}$ in column $P_2$. One can check that the accessed elements are sufficient to rebuild the two erased columns, and the rebuilding ratio is $2/3=e/r$. It can be shown that similar rebuilding can be done for any two systematic node erasures.  Therefore, in this example the lower bound of the rebuilding ratio is achievable.
\end{xmpl}

In the next Lemma, we consider an information array of size $p\times k$ and an $(n,k)$ MDS optimal-update code with $r=n-k$ parity nodes. Each parity node $l\in [0,r-1]$ is constructed from the set of permutations $\{f_i^l\}$ for $i \in [0,k-1]$. Notice that in the general case the number of rows $p$ in the array is not necessarily a power of $r$. From now on, we assume WLOG that columns $[0,e-1]$ are erased.

\begin{lem}\label{lem0120}
Below are sufficient and necessary conditions for optimal rebuilding ratio for $e$ erasures in an MDS optimal-update code:\\
There exists a set $X \subseteq [0,p-1]$ of size $|X|=ep/r$, such that
\begin{enumerate}
	\item For any parity node $l, l\in [0,r-1],$ the group $G^l$ {\bf stabilizes} the set $X$, i.e., for any $g\in G^l$,
	\begin{equation}
   		g(X)=X,
			\label{eq03171}
		\end{equation}
		where $G^l$ is generated by the set of permutations
		$\{f^{-l}_e\circ f^l_j\}_{j=e}^{k-1}.$
  \item For any erased column $i\in [0,e-1]$,
  	\begin{equation}
  		\cup_{l=0}^{r-1} (f_i^l)^{-1} f_e^l(X) = [0,p-1].
  		\label{eq03172}
		\end{equation}
  \item The $ep$ equations (zigzag sets) defined by the set $X$ are linearly independent.
\end{enumerate}
\end{lem}
\begin{IEEEproof}
In an erasure of  $e$ columns, $ep$ elements need rebuilt, hence we need $ep$ equations (zigzags) that contain these elements. In an optimal rebuilding scheme, each parity node contributes $ep/r$ equations by accessing the values of  $ep/r$ of its zigzag elements. Moreover, the union of the zigzag sets that create these zigzag elements constitutes an $e/r$ portion of the elements in the surviving systematic nodes. In other words, assume that we access rows $X$ from the surviving columns $[e,k-1]$, $X \subseteq [0,p-1]$, then $|X|=ep/r$ and
\begin{equation*}
f_j^l(X)=f_i^l(X)
\end{equation*}
for any parity node $l \in [0,r-1]$ and $i,j \in [e,k-1]$. Note that it is equivalent that for any
parity node $l \in [0,r-1]$ and surviving systematic node $j \in [e,k-1]$
\begin{equation*}
f_j^l(X)=f_e^l(X).
\end{equation*}
Let $G^l$ be the subgroup of the symmetric group $S_p$ that is generated by the set of permutations $\{f^{-l}_e\circ f^l_j\}_{j=e}^{k-1}$, where $f^{-l}_e=(f^{l}_e)^{-1}$ is the inverse of $f^{l}_e$. It is easy to see that the previous condition is also equivalent to that for any parity $l\in [0,r-1]$ the group $G^l$ {stabilizes} $X$, i.e., for any $f\in G^l,f(X)=X.$

Assuming there is a set $X$ that satisfies this condition, we want to rebuild the $ep$ elements from the chosen $ep$ equations, i.e., the $ep$ equations with the $ep$ variables being solvable. Firstly we need the condition that each element in the erased column will appear at least once in the chosen zigzag sets (equations). Parity $l\in [0,r-1]$ accesses its zigzag elements $f^l_e(X)$, and these zigzag sets contain the elements in rows $f_i^{-l} f_e^l(X)$  of the erased column $i\in [0,e-1]$. Hence the condition is equivalent to that for any erased column $i\in [0,e-1]$,
\begin{equation*}
\cup_{l=0}^{r-1} (f_i^l)^{-1} f_e^l(X) = [0,p-1].
\end{equation*}
In addition, we need to make sure that the $ep$ equations are linearly independent, which depends on the coefficients in the linear combinations that created the zigzag elements.
\end{IEEEproof}

Next we will interpret these conditions in the special case where the number of rows $p=r^m$, and the permutations are generated by $T=\{v_0,v_1,\dots,v_{k-1}\}$ $\subseteq$ $\mathbb{F}_r^m$ and Construction \ref{cnstr5}, i.e., $f_i^l(x)=x+l v_i$ for any $x \in [0,r^m-1]$.
From here on we will focus only on the special case with $p=r^m$ and $r$ prime. 
\begin{cor} \label{cor0120}
Below are sufficient and necessary conditions for optimal rebuilding ratio in $e$ erasures for general zigzag codes:\\
There exists a set $X \subseteq \mathbb{F}_r^m$ of size $|X|=er^{m-1}$, such that
\begin{enumerate}
	\item $X$ is a union of cosets of the subspace $$Z=\spun\{v_{e+1}-v_e,\dots,v_{k-1}-v_e\}.$$
	\item For any erased column $i\in [0,e-1]$,
  	\begin{equation}
  	\label{eq03241}
  		\cup_{l=0}^{r-1} (X+l(v_i-v_e))=\mathbb{F}_r^m.	
		\end{equation}
  \item The $er^m$ equations (zigzag sets) defined by the set $X$ are linearly independent.
\end{enumerate}
\end{cor}
\begin{IEEEproof}
Consider the general zigzag codes of Construction \ref{cnstr5}. Note that in the case of $r$ a prime $$G^1=G^2=...=G^{r-1},$$ and in that case we simply denote the group as $G$. Condition 2) follows directly from the definition of the permutation $f_{i}^l$, and condition 3) is identical to Lemma \ref{lem0120}. We only need to show the following for condition 1):\\
Define $X\subseteq \mathbb{F}_r^m$ and $G$ as above, then $G$ stabilizes $X$, if and only if $X$ is a union of cosets of the subspace
\begin{equation}
 \label{eq03242}
Z=\spun\{v_{e+1}-v_e,\dots,v_{k-1}-v_e\}.
\end{equation}
It is easy to check that any coset of $Z$ is stabilized by $G$, hence if $X$ is a union of cosets it is also a stabilized set. For the other direction let $x,y\in \mathbb{F}^m_r$ be two vectors in the same coset of $Z$, it is enough to show that if $x\in X$ then also $y\in X$. Since $y-x\in Z$ there exist $\alpha_1,...,\alpha_{k-1-e}\in [0,r-1]$ such that
$y-x=\sum_{i=1}^{k-1-e}\alpha_i(v_{e+i}-v_e).$ Since $f(X)=X$ for any $f\in G$ we get that $f(x)\in X$ for any $x\in X$ and  $f\in G$, hence
\begin{align*}y&=x+y-x\\
&=x+\sum_{i=1}^{k-1-e}\alpha_i(v_{e+i}-v_e)\\
&=f_e^{-\alpha_{k-1-e}}f_{k-1}^{\alpha_{k-1-e}}...f_e^{-\alpha_{1}}f_{e+1}^{\alpha_{1}}(x) \in X,
\end{align*}
where we used the fact that $f_e^{-\alpha_{k-1-e}}f_{k-1}^{\alpha_{k-1-e}}...f_e^{-\alpha_{1}}f_{e+1}^{\alpha_{1}}\in G$. So $y \in X$ and the result follows.
\end{IEEEproof}

\textbf{Remark:} For any set of vectors $S$ and $v,u\in S$, $$\spun\{S-v\}=\spun\{S-u\}.$$ Here $S-v=\{v_i-v|v_i\in S\}.$ Hence, the subspace $Z$ defined in the previous theorem does not depend on the choice of the vector $v_e$. We can equivalently replace $v_e$ with any $v_i$, $i \in [e,k-1]$.

The following theorem gives a simple equivalent condition for conditions $1),2)$ in Corollary \ref{cor0120}.
\begin{lem} \label{thm03311}
There exists a set $X \subseteq \mathbb{F}_r^m$ of size $|X|=er^{m-1}$ such that conditions $1),2)$ in Corollary \ref{cor0120} are satisfied if and only if
\begin{equation} \label{eq03312}
 v_i-v_e \notin Z
\end{equation}
for any erased column $i\in [0,e-1].$
\end{lem}
\begin{IEEEproof}
Assume conditions $1),2)$ are satisfied. If $v_i-v_e\in Z$ for some erased column $i\in [0,e-1]$ then by condition 2),
$X=\cup_{l=0}^{r-1} (X+l(v_i-v_e))=\mathbb{F}_r^m,	$
which is a contradiction to $X\subsetneq \mathbb{F}_r^m.$
On the other hand, If \eqref{eq03312} is true, then $v_i-v_e$ can be viewed as a permutation that acts on the cosets of $Z$. The number of cosets of $Z$ is $r^m/|Z|$ and this permutation (when it is written in cycle notation) contains  $r^{m-1}/|Z|$ cycles, each with length $r$. For each $i\in [0,e-1]$ choose $r^{m-1}/|Z|$ cosets of $Z$, one from each cycle of the permutation $v_i-v_e.$ In total $er^{m-1}/|Z|$ cosets are chosen for the $e$ erased nodes.
Let $X$ be the union of the cosets that were chosen. It is easy to see that $X$ satisfies condition $2)$. If $|X|<er^{m-1}$ (Since there might be cosets that were chosen more than once) add arbitrary $(er^{m-1}-|X|)/|Z|$ other cosets of $Z$, and also condition $1)$ is satisfied.
\end{IEEEproof}

In general, if \eqref{eq03312} is not satisfied, the code does not have an optimal rebuilding ratio.
However we can define the \emph{optimal subspace} as
\begin{equation}
Z=\spun\{v_i-v_e\}_{i\in I},
\label{eq04131}
\end{equation}
where we assume w.l.o.g. $e\in I$ and  $I\subseteq [e,k-1]$ is a maximal subset of surviving nodes that satisfies for any erased node $j\in [0,e-1], v_j-v_e\notin Z.$
Hence from now on we assume that $Z$ is defined by a subset of surviving nodes $I$. This set of surviving nodes will have  an optimal rebuilding ratio (see Corollary \ref{cor0401}), i.e., in the rebuilding of columns $[0,e-1]$, columns $I$ will access a portion of $e/r$ of their elements. The following theorem gives a sufficient condition for the $er^m$ equations defined by the set $X$ to be solvable linear equations.

\begin{lem}\label{thm0413}
Suppose that there exists a subspace $X_0$ that contains the optimal subspace $Z$ such that for any erased node $i\in [0,e-1]$
\begin{equation} \label{eq04132}
X_0\oplus \spun\{v_i-v_e\} = \mathbb{F}_r^m,
\end{equation}
where $\oplus$ denotes the direct sum of two subspaces,
then over a large enough field, the set $X$ defined as an union of some $e$ cosets of $X_0$ satisfies the conditions in Corollary \ref{cor0120} and the corresponding general zigzag code is MDS.
\end{lem}
\begin{IEEEproof}
Condition $1)$ is trivial.
Note that by \eqref{eq04132}, $l(v_i-v_e) \notin X_0$ for any $l\in [1,r-1]$ and $i\in [0,e-1]$, hence $\{X_0+l(v_i-v_e)\}_{l\in [0,r-1]}$ is the set of cosets of $X_0$. Let $X_j=X_0+j(v_i-v_e)$
be a coset of $X_0$ for some $i \in [0,e-1]$ and suppose $X_j \subset X$. Now let us check condition 2):
\begin{align}
\cup_{l=0}^{r-1}(X+l(v_i-v_e))& \supseteq \cup_{l=0}^{r-1}(X_j+l(v_i-v_e)) \nonumber\\
&= \cup_{l=0}^{r-1}(X_0+j(v_i-v_e)+l(v_i-v_e)) \nonumber \\
&= \cup_{l=0}^{r-1}(X_0+(j+l)(v_i-v_e)) \nonumber \\
& = \cup_{t=0}^{r-1}(X_0+t(v_i-v_e)) \label{05192}\\
&=\mathbb{F}_r^m. \label{05193}
\end{align}
\eqref{05192} holds since $j+l$ is computed mod $r$.
So condition $2)$ is satisfied. 

Next we prove condition $3)$ and that the code is MDS. Let every coefficient in the encoding matrices $P_j^i$ (which are generalized permutation matrices) be an indeterminate. 
There are $er^m$ unknowns and $er^m$ equations during the rebuilding of $e$ erasures. Writing the equations in a matrix form we get $AY=b$, where $A$ is an $er^m \times er^m$ matrix. $Y,b$ are vectors of length $er^m$, and $Y=(y_1,...,y_{er^m})^T$ is the unknown vector. The matrix $A=(a_{i,j})$ is defined as $a_{i,j}=x_{i,j}$ if the unknown $y_j$ appears in the $i$-th equation, otherwise $a_{i,j}=0$. Hence we can solve the equations if and only if there is assignment for the indetermediates $\{x_{i,j}\}$ in the matrix $A$ such that $\det(A)\neq 0$.
By \eqref{05193}, accessing rows corresponding to any coset $X_j$ will give us equations where each unknown appears exactly once. Since $X$ is a union of $e$ cosets, each unknown appears $e$ times in the equations. Thus each column in $A$ contains $e$ indeterminates. Moreover, each equation contains one unknown from each erased node, thus any row in $A$ contains $e$ indeterminates.
Then by Hall's Marriage Theorem \cite{Hall} we conclude that there exists a permutation $f$ on the integers $[1,er^m]$ such that $$\prod_{i=1}^{er^m}a_{i,f(i)}\neq 0.$$ Hence the polynomial $\det(A)$ when viewed as a symbolic polynomial, is not the zero polynomial, i.e., $$\det(A)=\sum_{f\in S_{er^m}}\sgn(f)\prod_{i=1}^{er^m}a_{i,f(i)}\neq 0.$$ By \cite{Alon} we conclude that there is an assignment from a field large enough for the indeterminates such that $\det(A)\neq 0$, and the equations are solvable. Note that this proof is for a specific set of erased nodes. However if \eqref{eq04132} is satisfied for any set of $e$ erasures, multiplication of all the nonzero polynomials $\det(A)$ derived for any set of erased nodes is again a nonzero polynomial. 
To show that the code MDS, we can use the remark after Theorem \ref{thm1}. We multiply the above nonzero polynomial by $\det(G')$ for all $G'$ as in \eqref{eqG}. Then we again obtain a nonzero polynomial. 
We conclude that there is an assignment over a field large enough such that any of the  matrices $A$ and $G'$ is invertible, and the result follows.
\end{IEEEproof}

Suppose the subspace $X_0$ in Lemma \ref{thm0413} exits for a subset of nodes $I$, we give an upper bound for the rebuilding ratio. Notice here that the ratio is only calculated for the specific erasures $[0,e-1]$ instead of averaging over all erasure cases.
\begin{cor} \label{cor0401}
Lemma \ref{thm0413} requires rebuilding ratio at most
$$\frac{e}{r}+\frac{(r-e)(k-|I|-e)}{r(k+r-e)}.$$
\end{cor}
\begin{IEEEproof}
By Lemma \ref{thm0413}, the fraction of accessed elements in columns $I$ and the parity columns is $e/r$ of each column. Moreover, the accessed elements in the rest columns are at most an entire column. Therefore, the rebuilding ratio is at most
$$\frac{\frac{e}{r}(|I|+r)+(k-|I|-e)}{k+r-e}
= \frac{e}{r}+\frac{(r-e)(k-|I|-e)}{r(k+r-e)}$$
and the result follows.
\end{IEEEproof}

Note that as expected when $|I|=k-e$ the rebuilding ratio is optimal, i.e., $e/r$.

In order to use Lemma \ref{thm0413}, we need to find a subspace $X_0$ as in \eqref{eq04132}. The following theorem shows that such a subspace always exists, moreover it gives an explicit construction of it.

\begin{lem}
\label{thm0414}
Suppose $1 \le e<r$ erasures occur.
Let $Z$ be defined by \eqref{eq04131} that satisfy Lemma \ref{thm03311}, namely, $v_i-v_e \notin Z$ for any erased node $i \in [0,e-1]$. Then there exists $u\perp Z$ such that for any $i \in [0,e-1]$,
\begin{equation}\label{eq04144}
u \cdot (v_i-v_e) \neq 0.
\end{equation}
Moreover the orthogonal subspace $X_0=(u)^\perp$ satisfies \eqref{eq04132} in Lemma \ref{thm0413}. 
\end{lem}

\begin{IEEEproof}
First we will show that such vector $u$ exists. Let $u_1,...u_t$ be a basis for $(Z)^\perp$, the orthogonal subspace of $Z.$
Any vector $u$ in $(Z)^\perp$ can be written as $u = \sum_{j=1}^t x_j u_j$ for some unknown $x_j$'s.
We claim that for any $i \in [0,e-1]$ there exists $j$ such that $u_j \cdot(v_i-v_e) \neq 0$. Because otherwise, $(Z)^{\perp}=\spun\{u_1,\dots,u_t\} \perp (v_i-v_e)$, which means $v_i-v_e \in Z$ and reaches a contradiction. Thus the number of solutions for $(x_1,\dots,x_t)$ in the linear equation
$$\sum_{j=1}^{t}x_ju_j \cdot (v_i-v_e)=0$$
is at most $r^{t-1}$, which equals the number of $u$ such that $u\cdot (v_i-v_e)=0$.
 Hence by the union bound there are at most $er^{t-1}$ vectors $u$ in $(Z)^\perp$ such that $u\cdot (v_i-v_e)=0$ for some erased node $i\in [0,e-1]$. Since $|(Z)^\perp|=r^t>er^{t-1}$ there exists $u$ in $(Z)^\perp$ such that for any erased node $i\in [0,e-1]$, $$u\cdot (v_i-v_e)\neq 0.$$ Define $X_0=(u)^\perp$, and note that for any erased node $i\in [0,e-1],v_i-v_e\notin X_0$, since $u\cdot (v_i-v_e)\neq 0$ and $X_0$ is the orthogonal subspace of $u$. Moreover, since $X_0$ is a hyperplane we conclude that
$$X_0\oplus \spun\{v_i-v_e\}=\mathbb{F}_r^m,$$
and the result follows.
\end{IEEEproof}

Lemmas \ref{thm0413} and \ref{thm0414} give us  an algorithm to rebuild multiple erasures.

\begin{algorithm} 
\begin{algorithmic}[1]
\State Find $Z$ by \eqref{eq04131} satisfying \eqref{eq03312}.
\State Find $u \perp Z$ satisfying \eqref{eq04144}. Define $X_0=(u)^{\perp}$ and $X$ as a union of $e$ cosets of $X_0$.
\State Access rows $f_e^l(X)$ in parity $l \in [0,r-1]$ and all the corresponding information elements.
\end{algorithmic}
\caption{Rebuilding algorithm for a zigzag code satisfying Lemma \ref{thm0413}.}
\label{alg0120} 
\end{algorithm}

{\bf Remark:} By Lemma \ref{thm0413} we know that under a proper selection of coefficients the rebuilding is possible for any $e$ erasures, $1 \le e \le r$, and the code can be made MDS.

In the following we give two examples of rebuilding using this algorithm.
The first example shows an optimal rebuilding for any set of $e$ node erasures. As mentioned above, the optimal rebuilding is achieved since \eqref{eq03312} is satisfied for all surviving nodes, i.e., $I=[e,k-1]$.
\begin{xmpl} \label{xmpl0331}
Let $T=\{v_0,v_1,\dots,v_m\}$ be a set of vectors that contains an orthonormal basis of $\mathbb{F}_r^m$ together with the zero vector. Suppose columns $[0,e-1]$ are erased. Note that in that case $I=[e,m]$ and $Z$ is defined as in \eqref{eq04131}. Depending on whether the zero vector is erased or not, we have two different cases. When $v_0=0$ define $$u=\sum_{j=e}^mv_j.$$ When $v_e=0$ define $$u=\sum_{j=0}^{e-1}v_j.$$
Let $X_0=(u)^\perp$.
It is easy to check that $u\perp Z$ and for any erased column $i\in [0,e-1],u\cdot (v_i-v_e)\neq 0$. Notice that $|I|=k-e$. Therefore by Lemma \ref{thm0414} and Corollary \ref{cor0401}, a set $X$ defined as a union of an arbitrary $e$ cosets of $X_0$ satisfies has optimal rebuilding ratio $e/r$.
Moreover, it can be verified that when $r=3$, the coefficients in \eqref{eq925} satisfies condition 3). Therefore, when $r=3$, we have an MDS code with $\mathbb{F}_4$ and optimal rebuilding for all $e \le 3$.
\end{xmpl}

In the case of a single erasure, Example \ref{xmpl0331} coincides with the rebuilding algorithm in Theorem \ref{thm_cnstr1}.
In the example of Figure \ref{fig5}, we know that the vectors generating the permutations are the standard basis (and thus are orthonormal basis) and the zero vector. When columns $C_0,C_1$ are erased, $u=e_2$ and $X_0=(u)^{\perp}=\spun\{e_1\}=\{0,3,6\}$. Take $X$ as the union of $X_0$ and its coset $\{1,4,7\}$, which is the same as Example \ref{xmpl0503}. One can check that each erased element appears exactly 3 times in the equations and the equations are solvable in $\mathbb{F}_4$. Similarly, the equations are solvable for other $2$ systematic erasures.

We summarize the result of the above example in the next theorem. In particular, when $T$ is the set of standard basis and the zero vector as in Theorem \ref{thm_cnstr1}, we obtain an optimal rebuilding code.
\begin{thm}\label{thm0120}
The optimal zigzag code is an MDS code with $n=m+r,k=m+1,p=r^m$, for any prime $r$ and positive integer $m$, that has optimal rebuilding ratio of $e/r$ for any set of $e$ erasures, $1 \le e \le r$.
\end{thm}

In the following example the code has $O(m^2)$ columns. The set $I$ does not contain all the surviving systematic nodes, hence the rebuilding is not optimal but is at most $\frac{1}{2}+O(\frac{1}{m})$.

\begin{xmpl}
 \label{xmpl0401}
Suppose $2|m$. Let $T=\{v=(v_1,\dots,v_m):\|v\|_1=2,v_i=1,v_j=1, \text{ for some } i \in [1,m/2], j \in [m/2+1,m]\subset \mathbb{F}_2^m$ be the set of vectors generating the code with $r=2$ parities, hence the number of systematic nodes is $|T|=k=m^2/4$. Suppose column $w=(w_1,\dots,w_m)$, $w_1=w_{m/2+1}=1$ is erased. Define the set $I=\{v\in T: v_1=0\},$ and
$$Z=\spun\{v_i-v_e|i\in I\}$$
for some $e \in I$. Thus $|I|=m(m-2)/4$.
It can be seen that $Z$ defined by the set $I$ satisfies \eqref{eq03312}, i.e., $w-v_e\notin Z$  since the first coordinate of a vector in $Z$ is always $0$, as oppose to $1$ for the vector $w-v_e$. Define $u=(0,1,...,1)$ and $X_0=(u)^\perp.$ It is easy to check that $u\perp Z$ and   $u\cdot (w-v_e)= 1\neq 0.$ Hence, the conditions in Lemma \ref{thm0414} are satisfied and rebuilding can be done using $X_0$.
Moreover by Corollary \ref{cor0401} the rebuilding ratio is at most
$$\frac{1}{2}+\frac{1}{2}\frac{(m/2)-1}{(m^2/4)+1} \approx \frac{1}{2}+\frac{1}{m}.$$
Note that by proper coefficients assignment (similar to Construction 3 in \cite{Tamo2}) we can use a field of size $5$ or $8$ to assure that the code is an MDS code.
\end{xmpl}

\subsection{Minimum Number of Erasures with Optimal Rebuilding}
Next we want to point out a surprising phenomena in terms of how many erasures can be optimally rebuilt for a given code.
We say that a set of vectors $S$ satisfies \emph{property} $e$ for $e\geq 1$ if for any subset $A\subseteq S $ of size $e$ and any $u\in A$,
$$u-v\notin \spun\{w-v:w\in S\backslash A\},$$
where $v\in S\backslash A$. Recall that by Lemma \ref{thm03311} any set of vectors that generates a code $\cC$ and can rebuild optimally any $e$ erasures should satisfy  property $e$. The following theorem shows that this property is monotonic, i.e., if $S$ satisfies property $e$ then it also satisfies property $a$ for any $e\leq a \leq |S|.$

\begin{thm}
Let $S$ be a set of vectors that satisfies property $e$, then it also satisfies  property $a$, for any $e \leq a \leq |S|$.
\end{thm}

\begin{IEEEproof}
Let $A\subseteq S,|A|=e+1$ and assume to the contrary that $u-v\in \spun\{w-v:w\in S\backslash A\}$ for some $u\in A$ and $v\in S\backslash A$. $|A|\geq 2$ hence there exists $x\in A\backslash \{u\}$. It is easy to verify that $u-v\in \spun\{w-v:w\in S\backslash A^*\}$, where $A^*=A\backslash \{x\}$ and $|A^*|=e$ which contradicts the property $e$ for the set $S$.
\end{IEEEproof}
Hence, from the previous theorem we conclude that a code $\cC$ that can rebuild optimally $e$ erasures, is able to rebuild optimally any number of erasures greater than $e$ as well. However, as pointed out  already there are codes with $r$ parities that can not rebuild optimally from some $e<r$ erasures. Therefore, one might expect to find a code $\cC$ with parameter $e^* \ge 1$ such that it can rebuild optimally $e$ erasures \emph{only} when $e^*\leq e\leq r$. For example, for $r=3,m=2$ let $\cC$ be the zigzag code constructed by the vectors $\{0,e_1,e_2,e_1+e_2\}$. It was proven in \cite{Tamo2} that if an MDS optimal-update code optimally rebuilds any single erasure, and has $p=r^m$, then
\begin{align} \label{eq0120}
k \le m+1.
\end{align}
Thus the above code $\cC$ cannot rebuild one erasure optimally.
However, one can check that for any two erased columns, the conditions in Lemma \ref{thm0413} are satisfied hence the code can rebuild optimally for any $e=2$ erasures and we conclude that  $e^*=2$ for this code.

The phenomena that some codes has a threshold parameter $e^*$, such that \emph{only} when the number of erasures $e$ is at least the threshold $e^*$ can the code rebuild optimally, is counter intuitive and surprising. This phenomena gives rise to another question. 
From \eqref{eq0120} a code with optimal rebuilding for single erasure must have $k \le m+1$.
Can $k$ be increased in a code with an optimal rebuilding of $e$ erasures, $e > 1$? The previous example answers this question affirmatively and shows a code with $k=4=m+2$ can rebuild optimally any $e=2$ erasures, but not 1 erasure.
The following theorem gives an upper bound for the maximum systematic columns in a code that rebuilds optimally any $e$ erasures, and it coincides with \eqref{eq0120} when $e=1$.

\begin{thm}
Let $\cC$ be a zigzag code constructed by Construction \ref{cnstr5} using vectors from $\mathbb{F}_r^m$. If $\cC$ can rebuild optimally any $e$ erasures, for some $1\le e<r$, then  the number of systematic columns $k$ in the code satisfies
$$k \le m+e.$$
\end{thm}
\begin{IEEEproof}
Consider a code with length $k$ and generated by vectors $v_0,v_1,\dots,v_{k-1}.$ If these vectors are linearly independent then $k\leq m$ and we are done. Otherwise they are dependent.
Suppose $e$ columns are erased,  $1 \le e<r$. Let $v_{e}$ be a surviving column. Consider the set a of vectors: $R=\{v_i-v_{e}: i \in [0,k-1],i\neq e\}.$
We know that the code can rebuild optimally only if \eqref{eq03312} is satisfied for all possible $e$ erasures. Thus for any $i \neq e$, $i \in [0,k-1],$ if column $i$ is erased and column $e$ is not, we have
$v_i-v_e \notin Z$ and thus $v_i-v_e \neq 0$.
So every vector in $R$ is nonzero.
Let $s$ be the minimum number of dependent vectors in $R$, that is, the minimum number of vectors in $R$ such that they are dependent.
For nonzero vectors, we have $s \ge 2$.
Say $\{v_{e+1}-v_{e},v_{e+2}-v_e,\dots,v_{e+s}-v_e\}$ is a minimum dependent set of vector. Since any $m+1$ vectors are dependent in $\mathbb{F}_r^m$,
$$s \le m+1.$$
We are going to show $k-e \le s-1$.
Suppose to the contrary that the number of remaining columns satisfies $k-e \ge s$ and $e$ erasures occur. When column $v_{e+s}$ is erased and the $s$ columns $\{v_e,v_{e+1},\dots,v_{e+s-1}\}$ are not, we should be able to rebuild optimally.
However since we chose a dependent set of vectors, $v_{e+s}-v_e$ is a linear combination of $\{v_{e+1}-v_{e},v_{e+2}-v_e,\dots,v_{e+s-1}-v_e\}$, whose span is contained in $Z$ in \eqref{eq03312}. Hence \eqref{eq03312} is violated and we reach a contradiction. Therefore,
$$k-e \le s-1 \le m.$$
\end{IEEEproof}

We know that this upper bound is tight in some cases. For $e=1$ we already gave codes with optimal rebuilding of $1$ erasure and $k=m+1$ systematic columns in Theorem \ref{thm_cnstr1}. Moreover, for $e=2$ the code constructed by the vectors $\{0,e_1,e_2,e_1+e_2\}$ reaches the upper bound with $k=4$ systematic columns. It is an open research problem whether this bound is achievable for any $k,m,e$.

\section{Optimal Rebuilding for Any Node}
\label{sec5}
%
Zigzag code in Theorem \ref{thm_cnstr1} has optimal rebuilding for systematic nodes. However, in order to rebuild a parity node, one has to access all the information elements.
In this section we construct MDS codes with optimal rebuilding ratio for a single systematic or parity node. The code has $k=m-1$ systematic nodes, $r$ parities nodes, and $r^{m}$ rows, for any $m,r$.

We first describe the code construction using encoding matrices, and then prove its optimality for rebuilding, and at last show its MDS property.

\subsection{Code Construction}

For $i \in [0,r-1]$, define $X_i$ as the $i$-th set of vectors of size $r^{m}$ lexicographically, namely, $$X_i=\{v \in \mathbb{Z}_r^m: v \cdot e_1 = i\}.$$
$X_0$ is a subgroup of $\mathbb{Z}_r^m$.
Define for $j \in [2,m]$ a permutation $f_j(x)=x+e_j+e_1$, $x \in [0,r^{m}-1]$. So the permutation matrix $P_j$ corresponding to $f_j$ can be written as a $r \times r$ block matrix:
\begin{equation}
\label{eq1}
P_j=\left( \begin{matrix}
 & & & p_j \cr
 p_j & & & \cr
 & \ddots & & \cr
& & p_j &
\end{matrix} \right),
\end{equation}
where $p_j$ of size $r^{m-1} \times r^{m-1}$ corresponds to the restricted mapping of $f_j:X_i \mapsto X_{i+1}$. In particular, represent integer $l,i \in [0,r^{m-1}-1]$ by a $r$-ary vector of length $m$ by appending a $0$ in the first coordinate. We can view $l,i$ as vectors in $X_0$. Then the $(i,l)$-th entry of $p_j$ is 1 if and only if $$l+e_j=i,$$ for $i,l \in X_0$. 
It is clear that $p_j$ is a matrix of order $r$, i.e., $p_j^r$ is the identity permutation and $p_j^l \neq 1$ for any $0 < l < r$. 
In this section, the superscript of $p_j$ are computed modulo $r$. For example, $p_j^{i-i'}$ denotes $p_j^{i-i' \mod r}$, for integers $i,i'$.

Next we assign coefficients in $P_j$.
In order to ensure MDS property, we modify $p_j$ to generalized permutation matrices the same way as in zigzag code \eqref{eq925}\eqref{eq926}, $j \in [2,m]$. Denote by $p_j(i,l)$ the $(i,l)$-th element of $p_j$.
When $r=2,3$, 
\begin{align}\label{eq21}
  p_j(i,l) = \begin{cases}
  c, & \textrm{if } l \cdot \sum_{t=2}^{j} e_t=0, l+e_j = i,\\
  1, & \textrm{if } l \cdot \sum_{t=2}^{j} e_t\neq 0, l+e_j = i,\\
  0, & \textrm{o.w.}
  \end{cases}
\end{align}
Note here the summation $\sum_{t=2}^{j} e_t$ is from $t=2$ since we will index the systematic nodes by $\{2,3,\dots,m\}$.
Here $c$ is  a primitive element in $\mathbb{F}_3,\mathbb{F}_4$, for $r=2,3,$ respectively. When $r \ge 4$, 
\begin{align}\label{eq22}
  p_j(i,l) = \begin{cases}
  \lambda_j, & \textrm{if } l+e_j = i,\\
  0, & \textrm{o.w.}
  \end{cases}
\end{align}

In the following, we will use block matrices the same as single elements. When referring to row or column indices, we mean block row or column indices. We refer to $p_j$ as a small block, and the corresponding block row or column as a small block row or column. And $P_j$ is called a big block with big block row or column.

\begin{figure}
  \centering
  \includegraphics[width=.4\textwidth]{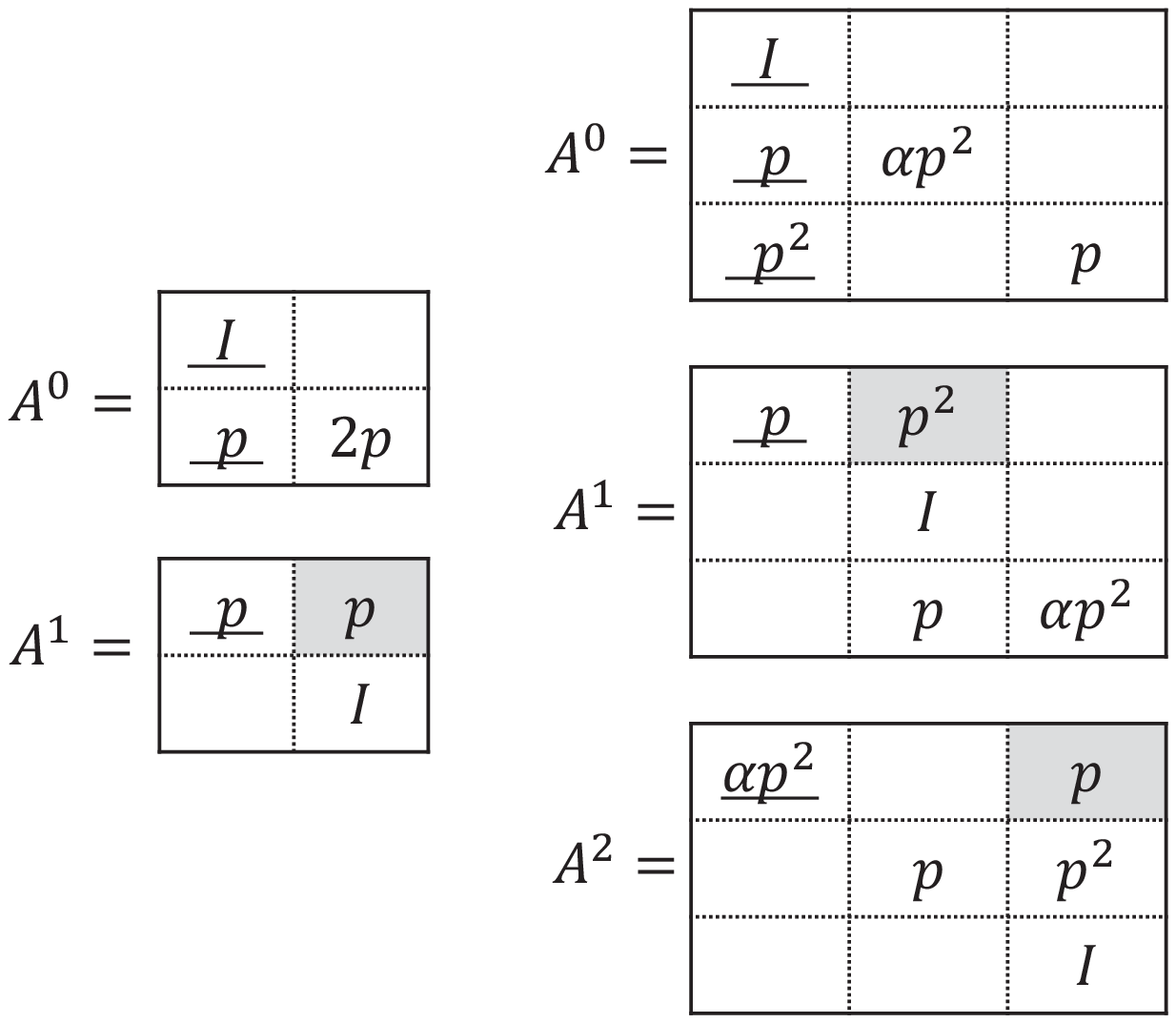}
  \caption{Parity matrices $A^i$ for $r=2$ (left) and $r=3$ (right) parities. When the 0th parity node is erased, the 0th row block $X_0$ is accessed from very surviving node. The underlined elements are computed from the accessed information elements. The remaining unknown terms in $A^0$ are recovered by the shaded elements from parity nodes.}
  \label{fig2-p}
\end{figure}

To define the encoding matrices, we first define some subset of indices $L_i$, for any $i \in [0,r-1]$:
$$L_i =
\begin{cases}
  \{i+1,\dots,i+ \frac{r-1}{2}\}, & r \textrm{ odd}, \\
 \{i+1,\dots,i+ \frac{r}{2}\}, & r \textrm{ even}, 0 \le i < \frac{r}{2},\\ 
  \{i+1,\dots,i+ \frac{r}{2}-1\}, & r \textrm{ even}, \frac{r}{2} \le i \le r-1,
\end{cases}
$$
where additions and subtractions are modulo $r$.
Define its complement excluding $i$ as $\overline{L_i} = \{0,1,\dots,i-1,i+1,\dots,r-1\}\backslash L_i$. Note that $i' \in L_i$ if and only if $i \in \overline{L_{i'}}$.

Let us define the encoding matrix $A_j^i$ for the $j$-th systematic node and the $i$-th parity, $j \in [2,m], i \in [0,r-1]$.
Let $A_j^i(x,y)$ be the $r^{m-1} \times r^{m-1}$ small block in block row $x$ and block column $y$ of the big block $A_j^i$, $x,y \in [0,r-1]$. Then
$$ A_j^i(x,y)=
\begin{cases}
  p_j^{-i+x}, &  y=i, 0 \le x \le r-1, \\
  \alpha p_j^{i-x}, & x=y \in L_i, \\
  p_j^{i-x}, & x=y \in \overline{L_i},\\
  0, & \textrm{o.w.} 
\end{cases}
$$ 
where $\alpha \neq 0,1$ is an element of the finite field $\mathbb{F}$.
Therefore, for $i=0,j\in [2,m]$, we have
  $$A_j^0=
  \bordermatrix{
     ~ & 0 & 1 & 2 & \cdots & r-2 & r-1 \cr
     0 & I & & & & & \cr
     1 & p_j  & \alpha p_j^{r-1}& & & &\cr
     2 & p_j^2 & & \alpha p_j^{r-2} & & &\cr
     \vdots & \vdots & & & \ddots & &\cr
     r-2 & p_j^{r-2} & & & & p_j^{2} & \cr
     r-1 & p_j^{r-1} & & & & & p_j}
  $$
  where $\alpha $ is multiplied to the diagonal in rows $L_0=\{1,\dots,\lfloor\frac{r}{2}\rfloor\}$.
  Moreover, $A_j^i$ is obtained by cyclicly shifting the rows and columns of $A_j^0$ to the right and bottom by $i$ positions (with possibly one change in a coefficient):
  $$A_j^i=
  \bordermatrix{
   ~ & 0 & \cdots & i-1 & i & i+1 & \cdots & r-1 \cr
   0 & \beta_1 p_j^i & & & p_j^{r-i} & & &\cr
   \vdots & & \ddots & & \vdots & & & \cr
   i-1 & & & \beta_{i-1} p_j & p_j^{r-1} & & & \cr
   i & & & & I & & & \cr
   i+1 & & & & p_j & \beta_{i+1} p_j^{r-1} & &\cr
   \vdots & & & & \vdots & & \ddots & \cr
   r-1 & & & & p_j^{r-i-1} & &  & \beta_{r-1} p_j^{i+1}
  }.
  $$
  where 
  $$\beta_x = \begin{cases}
  \alpha, & \textrm{if } x \in L_i,\\
  1, & \textrm{if } x \in \overline{L_i}.
  \end{cases}
  $$

\begin{cnstr}
  \label{cnstr2}
  Let $k=m-1$. Suppose the information array is of size $r^m \times k$, and the systematic nodes are indexed $[2,m]$, the parity nodes are indexed $[0,r-1]$.
 Let the first $k$ nodes be systematic, and the last $r$ nodes be parities. 
 The generator matrix of the code is
  $$\left(
  \begin{array}{ccc}
    I & & \\
    & \ddots & \\
    & & I \\
    A_2^0 & \cdots & A_{m}^0 \\
    \vdots & & \vdots \\
    A_2^{r-1} & \cdots & A_{m}^{r-1}
  \end{array}
  \right).$$
\end{cnstr}
{\bf Notation:} It should be noted that the superscript in $A_j^i$ does not denote power of the matrix, but the superscript in $p_j^i$ does denote the power of $p_j$. 
Sometimes we will omit the subscript $j$ when it is clear in the context.

\begin{figure}
  \centering
  \begin{tabular}{|c|c|c||c|c|}
\hline
~ & $C_2$  & $C_3$ & $P_0$ & $P_1$ \\
\hline
0 &$ a_{0,1}$ &$ a_{0,2} $& $a_{0,1}+a_{0,2} $ & $a_{2,1}+a_{6,1}+a_{1,2}+a_{5,2}$ \\
\hline
1 &$ a_{1,1}$ &$ a_{1,2} $& $a_{1,1}+a_{1,2} $ & $a_{3,1}+a_{7,1}+2a_{0,2}+2a_{4,2}$ \\
\hline
2 &$ a_{2,1}$ &$ a_{2,2} $& $a_{2,1}+a_{2,2} $ & $2a_{0,1}+2a_{4,1}+2a_{3,2}+2a_{7,2}$ \\
\hline
3 &$ a_{3,1}$ &$ a_{3,2} $& $a_{3,1}+a_{3,2} $ & $2a_{1,1}+2a_{5,1}+a_{2,2}+a_{6,2}$ \\
\hline
4 &$ a_{4,1}$ &$ a_{4,2} $& $a_{2,1}+2a_{6,1}+a_{1,2}+2a_{5,2}$ & $a_{4,1}+a_{4,2} $ \\
\hline
5 &$ a_{5,1}$ &$ a_{5,2} $& $a_{3,1}+2a_{7,1}+2a_{0,2}+a_{4,2}$ & $a_{5,1}+a_{5,2} $ \\
\hline
6 &$ a_{6,1}$ &$ a_{6,2} $& $2a_{0,1}+a_{4,1}+2a_{3,2}+a_{7,2}$ & $a_{6,1}+a_{6,2} $ \\
\hline
7 &$ a_{7,1}$ &$ a_{7,2} $& $2a_{1,1}+a_{5,1}+a_{2,2}+2a_{6,2}$ & $a_{7,1}+a_{7,2} $ \\
  \hline
  \end{tabular}
  \caption{An MDS array code with two systematic and two parity nodes by Construction \ref{cnstr2}. The finite field used is $\mathbb{F}_3$. }
  \label{fig3}
\end{figure}

\begin{xmpl}
\label{xmpl1}
  For codes with two and three parities, the encoding matrices $A^i$ are shown in Figure \ref{fig2-p}. When $r=2$, as finite field $\mathbb{F}_3$ is used, we can take $\alpha=2 \neq 1$. Coefficient $\alpha=2$ is multiplied to only the second diagonal in $A^0$.
  When $r=3$, finite field $\mathbb{F}_4$ is used and we choose some element $\alpha \neq 0,1, \alpha \in \mathbb{F}_4$. 
  It can be seen that $A^1,A^2$ are simply shifted versions of $A^0$. An example of a code with $r=2,m=3,k=2$, is shown in Figure \ref{fig3}.
\end{xmpl}

It can be seen from Construction \ref{cnstr2} and Figure \ref{fig3} that this code is not optimal update. In fact each information element appears $2r-1$ times in the parities.

\subsection{Optimal Rebuilding for Systematic and Parity Nodes}

Next we show that the code in Construction \ref{cnstr2} has optimal rebuilding ratio.
We first make some observations which are the key in proving the optimality of the rebuilding schemes. 
\begin{itemize}
  \item[(O1)] In small block row $i$ of $A^i = A_j^i$, there is only one non-zero small block, $I$, which is in block column $i$ or the diagonal entry. Therefore, for any row $l \in X_i$, the corresponding non-zero element is $1$ in column $l$ of $A^i$.
  \item[(O2)]
   In block row $i'$ of $A^i$, $i' \in L_i$, there are two non-zero small blocks, $p^{i'-i}, p^{i-i'}$, corresponding to block columns $i,i'$, respectively.
   Thus for every row (or parity) in this block row $i'$, there are two non-zero contributing terms.   
   Similarly, block row $i$ of $A_i'$ has two non-zero small blocks, $ \alpha p^{i'-i},  p^{i-i'}$, also corresponding to block columns $i,i'$, respectively. Writing these blocks together, we have:
    \begin{equation}
	\label{eq2}
	\bordermatrix{
	 ~  & &         i        &       &   i'      & \cr
	  i' \text{ in } A^i &  \cdots & p^{i'-i} & \cdots & \alpha p^{i-i'} & \cdots \cr
	   i \text{ in } A^{i'}~ & \cdots & p^{i'-i} & \cdots &  p^{i-i'} & \cdots
	}.
	\end{equation} 
    Moreover, if we consider row $l \in X_{i'}$ of $A^i$ and row $l+(i-i')e_1 \in X_{i}$ of $A^{i'}$, for $i' \in L_i$, we can see that they correspond to the the same columns (variables) but linearly independent terms for the parities $i,i'$:
    $$  b x_{l+ (i-i') (e_j+e_1)}+ \alpha c x_{l-(i-i') e_j},$$
    $$  b x_{l+ (i-i') (e_j+e_1)}+ c x_{l- (i-i') e_j},$$
    where $(x_0,\dots,x_{2^m})$ are the elements (variables) in systematic node $j$, $b$ is the  $(l-i'e_1, l-i'e_1-(i'-i)e_j)$-th entry in $p^{i'-i}$, and $c$ is the $(l-i'e_1, l-i'e_1-(i-i')e_j)$-th coefficient in $p^{i-i'}$.
\end{itemize}


%
%

\begin{thm}
  The code in Construction \ref{cnstr2} has optimal rebuilding ratio $1/r$ for rebuilding any node. More specifically, when the systematic node $e_i$ is erased, $i \in [2,m]$, we only need to access elements $Y_i=\{v \in \mathbb{Z}_r^m: v \cdot e_i=0\}$. When the parity $i$ is erased, $i \in [0,r-1]$, we only need to access elements $X_i=\{v \in \mathbb{Z}_r^m: v \cdot e_1 =i\}$.
\end{thm}
\begin{IEEEproof}
  {\bf Systematic rebuilding:} W.l.o.g. assume that column $e_2$ is erased. Access elements (equations) $Y:=Y_2=\{v \in \mathbb{Z}_r^m: v \cdot e_2=0\}$ from each parity.
  Let the elements in information node $2$ be unknowns $(x_0,\dots,x_{r^{m}-1})$.  We treat elements in the remaining systematic columns $3,\dots,m$ as known constants, and consider each accessed parity element as an equation. So we will focus on the matrices $ A_2^i$, $i \in [0,r-1]$, which are coefficients of the unknowns in the equation.
  We will show that all the unknowns $(x_0,\dots,x_{r^m-1})$ in column $2$ are solvable from the accessed equations. First notice that $Y$ is a subgroup of $\mathbb{Z}_r^m$, and coset $Y-t e_1 = Y$ for any $t \in [0,r-1]$. So one can verify that elements in $\mathbb{Z}_r^m$ can be written as one of the following three cases:
  $$\mathbb{Z}_r^m=\{l,l-te_2,l+t(e_2+e_1): l \in Y, t \in \overline{L_0}\}.$$
  So we need to show that an unknown element indexed by these three cases is solvable.

  For any $l\in Y$, assume $l \in Y \cap X_{i'}$ for some $i'$. First, consider the accessed equation $l$ of parity $i'$.
  by (O1) $x_l$ is solvable from equation
  $$x_l=f$$
  for some constant $f$ computed from parity and surviving information elements.
  Next consider the pair of unknowns $(x_{l-t e_2}, x_{l+t(e_1 + e_2)})$ with $t=i-i', t \in \overline{L_0}$, i.e., $i' \in L_i$.
  We consider accessed equation $l \in Y \cap X_i'$ of parity $i$ and accessed equation $l+(i-i')e_1 \in Y \cap X_i$ of parity ${i'}$. 
  By (O2), we have equations
  \begin{eqnarray*}
  b x_{l+t (e_2+e_1)}+\alpha c x_{l-t e_2}&=&g \\
  b x_{l+t (e_2+e_1)}+ c x_{l-t e_2}&=&h
  \end{eqnarray*}
  for some coefficients $\alpha \neq 0,1$, $b,c \neq 0$, and constants $g,h$ computed from parity and the surviving elements. These equations are obviously independent.
  Hence all unknowns are solvable.

  Next we show that only elements in $Y$ are accessed for every surviving node, and thus the rebuilding ratio is $1/r$. For any parity node $i$, only rows $Y$ are accessed. The accessed elements in systematic node $j$ are indexed by the columns corresponding to the non-zero entries of $A_j^i$ in rows (equations) $Y$, $j \in [3,m],i \in [0,r-1]$.
  For a surviving systematic node $j$ and parity $i$,  $j \in [3,m], i \in [0,r-1]$, 
  we can see from (O1) that any element $l \in Y \cap X_{i}$ of parity $i$ corresponds to columns (or accessed systematic element) $l$ of $A_j^{i}$; for $i' \neq i$, from (O2) any element $l \in Y \cap X_{i'}$ of parity $i$ corresponds to columns $l+(i-i')(e_1+e_j)$ and $l-(i-i')e_j$ of $A_j^i$, both of which belong to $Y$.
  Thus only elements $Y$ are accessed from each node.

  {\bf Parity rebuilding:} Since the parities are all symmetric, w.l.o.g. suppose the $0$-th parity is erased. Access $X_0$ from each node. 
  Need to show this is sufficient to recover
  $$[A_2^0,\dots,A_{m}^0]C,$$
  where $C=[C_2,\dots,C_m]^T \in \mathbb{F}^{(m-1)2^m}$ is the vector of systematic elements. 
  In (O2) take $i'=0$, then from a surviving parity $i$ we can access block row $0$ of matrix $[A_2^{i},\dots,A_m^{i}]C$, $i \in [1,r-1]$:
  $$[\underbrace{\underline{\beta p_2^{i}} \cdots p_2^{-i} \cdots}_{A_2^{i}} 
  \underbrace{\underline{\beta p_3^{i}} \cdots p_3^{-i} \cdots}_{A_3^{i}} 
  \cdots
  \underbrace{\underline{\beta p_m^{i}} \cdots p_m^{-i} \cdots}_{A_m^{i}} ]C,$$
  where $\beta$ is 
  \begin{equation*}
  \beta =
  \begin{cases}
    1, & i \in L_0 \\
    \alpha, & i \in \overline{L_0}.
  \end{cases}
  \end{equation*}
  Since elements $X_0$ are accessed from the systematic nodes, the $0$-th block column in each $A_j^{i}$ corresponds to the accessed information elements, and can be subtracted from the parities. 
  They are marked with underlines, and thus we know the value of:
  $$[\underbrace{0 \cdots p_2^{-i} \cdots}_{A_2^{i}}
  \underbrace{0 \cdots p_3^{-i} \cdots}_{A_3^{i}}
  \cdots
  \underbrace{0 \cdots p_m^{-i} \cdots}_{A_m^{i}} ]C,$$  
  Multiplying this row by $\gamma = \alpha/\beta$, we get
  $$[\underbrace{0 \cdots \gamma p_2^{-i} \cdots}_{A_2^{i}} 
  \underbrace{0 \cdots \gamma p_3^{-i} \cdots}_{A_3^{i}} 
  \cdots 
  \underbrace{0 \cdots \gamma p_m^{-i}\cdots}_{A_m^{i}}]C.$$
  By adding back elements in $X_0$ of the systematic nodes with appropriate coefficients, we can rebuild the $i$-th row of $A^0$, for all $1 \le i \le r-1$:
  $$[\underbrace{\underline{p_2^{i}} \cdots \gamma p_2^{-i} \cdots}_{A_2^{0}}
   \underbrace{\underline{p_3^{i}} \cdots \gamma p_3^{-i} \cdots}_{A_3^{0}} 
   \cdots 
   \underbrace{\underline{p_{m}^{i}} \cdots \gamma p_m^{-i}\cdots}_{A_m^{0}}]C,$$
  where again the underlined elements marks the $0$-th block column in each $A_j^0$. 
  The $0$-th row in $A^0$ is
  $$[\underbrace{\underline{I} \cdots}_{A_2^{0}} \underbrace{\underline{I} \cdots}_{A_3^{0}} \cdots \underbrace{\underline{I}\cdots}_{A_m^{0}}]$$
  and can be rebuilt from elements $X_0$ of the systematic nodes directly. Thus the erased node is rebuilt by accessing elements $X_0$ in every surviving node, which is a portion of $1/r$ of the elements.
\end{IEEEproof}

It can be seen from the above proof that the rebuilding of any single erasure can be easily implemented. If a systematic node is eared, we only need to solve at most two linear equations at a time, and the computation can be done in parallel. If a parity is erased the rebuilding is even simpler: we only need to subtract information elements with appropriate coefficients from the parities. Besides, the rebuilding above is different from Theorem \ref{thm_cnstr1}, where only one linear equation is solved at a time.

\begin{xmpl}
  \label{xmpl2}
  Consider the code with 2 or 3 parities in Figure \ref{fig2-p}. When the 0th parity node is erased, one can access elements $X_0$ from every surviving node, and therefore the underlined terms in the parities are known. Hence the sum of the shaded terms are known from the accessed parity elements. Adding the shaded terms and the underlined terms in $A^0$, we can rebuild parity $0$.
  
  For the example of Figure \ref{fig3}, when the systematic node $C_2$ is erased, one can access elements $Y_2=\{v: v \cdot e_2=0\}=\{0,1,4,5\}$ from all the surviving nodes. When the parity node $P_0$ is erased, one can access elements $X_0=\{0,1,2,3\}$ from all the remaining nodes. Then it is easy to check that in both cases it is sufficient to rebuild the erased column.
\end{xmpl}

\subsection{MDS Property}
Next we show the construction is indeed an MDS code.
Recall that we assigned the coefficients in the matrix $p_j$ as in \eqref{eq21}. And for $r=2,3$ parities, the assignment is as in \eqref{eq22}.
We will prove that if the coefficient assignment for the code in  Theorem \ref{thm_cnstr1} is MDS, then the code in Construction \ref{cnstr2} is also MDS. As a result, we do not need to design new coefficients for the construction, but simply reuse the encoding matrices (generalized permutation matrices) for the code in Theorem \ref{thm_cnstr1}.
First we make an observation on the small blocks.


\begin{lem}
\label{lem1}
  There exist coefficients $\lambda_j$ as in \eqref{eq21}, $j \in [2,m]$, such that any $t \times t$ sub-block matrix of
  \begin{align}\label{eq24}
  H'=\left(\begin{matrix}
    p_2^0 & \cdots & p_m^0 \\
    \vdots & & \vdots \\
    p_2^{r-1} & \cdots & p_m^{r-1}
  \end{matrix}\right)_{k \times k}
  \end{align}
  is invertible, for all $t \in [1,r]$. When $r=2,3$, the assignment in \eqref{eq22} satisfies the above condition.
\end{lem}
\begin{IEEEproof}
  Notice that $\{p_j^i\}$, $j \in [2,m],i \in [0,r-2]$, are the encoding matrices of Construction \ref{cnstr5} of information array size $r^{m-1} \times m$, shortened by deleting columns 0 and 1. Moreover all $t \times t$ sub-matrix of \eqref{eq24} being invertible, $1 \le t \le r$ is equivalent to Construction \ref{cnstr5} being MDS.
  Thus by Theorem \ref{thm1} the lemma holds.
\end{IEEEproof}

Next we show that Construction \ref{cnstr2} is MDS by computing the determinant of the matrices corresponding to different erasure patterns. Combining the result for the sub-block matrices in Lemma \ref{lem1}, the determinant can be shown to be non-zero.

\begin{figure}
\centering
\includegraphics[width=0.3\textwidth]{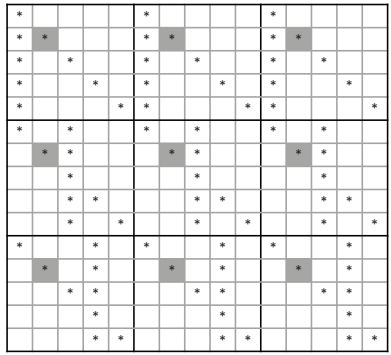}
\caption{Illustration of the matrix $A$ and its sub-matrices. The figure shows an example of $A$ with $r=5,t=3$ and $I=\{0,2,3\}$. Every square represents a small block matrix. The ``*'' marks non-zero sub-block matrices, and the other entries are all zero block matrices. The shaded sub-matrix is $A_x$ where $x=1$ is an index in $\bar{I}$. One can see that in column $x$ of every big block, there is only one non-zero entry.}
\label{fig:thmMDS}
\end{figure}

\begin{thm} \label{thmMDS}
There exist coefficients $\lambda_j$ as in \eqref{eq21}, $j \in [2,m]$, such that Construction \ref{cnstr2} is MDS.
When $r=2,3$, finite fields $\mathbb{F}_3, \mathbb{F}_4$, respectively, suffice for it to be MDS.
\end{thm}
\begin{IEEEproof}
Construction \ref{cnstr2} being MDS is equivalent to all of the following matrix being invertible:
  $$A=\left(\begin{matrix}
    A_{j_1}^{i_1} & \cdots & A_{j_t}^{i_1} \\
    \vdots & & \vdots \\
    A_{j_1}^{i_t} & \cdots & A_{j_t}^{i_t}
  \end{matrix}\right)_{t \times t} ,$$
  where each big block $A_j^i$ is of size $r^m \times r^m$ and $t \in [1,r], I=\{i_1,\dots,i_t\} \subseteq [0,r-1], \{j_1,\dots,j_t\} \subseteq [2,m]$. Let the complement of $I$ be $\overline{I}=[0,r-1] \backslash I$. We proceed the proof in two steps. At the first step, in each big block consider the small block column $x$, for some $x \in \overline{I}$. The only non-zero small blocks in these columns are in small block rows $x$. See for example Figure \ref{fig:thmMDS}. Thus by Laplace expansion $\det(A)=s\det(A_x)\det(A_{\bar{x}})$, where $s=1$ or $-1$, $A_x$ is the submatrix of $A$ with small block rows and columns $x$ in each big block, and $A_{\bar{x}}$ is the submatrix of $A$ corresponding to the remaining rows and columns. Moreover,
  $$A_x=\left(
  \begin{matrix}
  \beta_1 p_{j_1}^{i_1-x} & \cdots & \beta_1 p_{j_t}^{i_1-x} \\
  \vdots & & \vdots \\
  \beta_t p_{j_1}^{i_t-x} & \cdots & \beta_t p_{j_t}^{i_t-x} \\
  \end{matrix}
  \right),$$
  where $\beta_1,\dots,\beta_t$ are $1$ or $\alpha$. But by Lemma \ref{lem1}, the above matrix can be made invertible. So we only need to look at the remaining submatrix $A_{\bar{x}}$. Again, we can take out another small block column and row from an index in $\overline{I}$ from each big block, and it is invertible by Lemma \ref{lem1}. Continue this process, we are left with only columns and rows of $I$ in each big block.

At the second step, for all $i,i' \in I, i' \in L_i$, consider row $i'$ in $A^i$ and row $i$ in $A^{i'}$. They are shown in \eqref{eq2}. One can do a sequence of elementary row operations and keep the invertibility of the matrix, and get
  $$
    \bordermatrix{
        &        & \hspace*{-0.3cm} i & \hspace*{-0.3cm}        & \hspace*{-0.3cm}  i'      &\hspace*{-0.3cm}        &  \hspace*{-0.3cm}       & \hspace*{-0.3cm} i  &    \hspace*{-0.3cm}     & \hspace*{-.5cm}  i'             &     \hspace*{-.5cm}    \cr
   i' \text{ in } A^i& \cdots & \hspace*{-0.3cm}0 & \hspace*{-0.3cm}\cdots & \hspace*{-0.3cm} p_{j_1}^{i-i'} &\hspace*{-0.3cm} \cdots &\hspace*{-0.3cm} \cdots &\hspace*{-0.3cm} 0  & \hspace*{-0.3cm} \cdots & \hspace*{-0.3cm} p_{j_t}^{i-i'} & \hspace*{-0.3cm}\cdots \cr
    i  \text{ in } A^{i'}& \cdots & p_{j_1}^{i'-i} & \hspace*{-0.3cm}\cdots &\hspace*{-0.3cm} 0 &\hspace*{-0.3cm} \cdots & \hspace*{-0.3cm}\cdots & \hspace*{-0.3cm}p_{j_t}^{i'-i} & \hspace*{-0.3cm}\cdots &\hspace*{-0.3cm} 0 & \hspace*{-0.3cm}\cdots}.
  $$
  Proceed this for all $i,i' \in I, i' \in L_i$, we are left with block diagonal matrix in each big block. 
Then similar to the first step, we are only need to consider sub-matrices like $A_x$, which are invertible by Lemma \ref{lem1}. Thus Construction \ref{cnstr2} is MDS. 
\end{IEEEproof}

For example, one can easily check that the code in Figure \ref{fig3} is able to recover the information from any two nodes. Therefore it is an MDS code.
Theorem \ref{thmMDS} implies that once we have an MDS code in Theorem \ref{thm_cnstr1}, we can use its coefficients and design a new code by Construction \ref{cnstr2}. And the new code is guaranteed to be an MDS code.

%
\section{Correcting Errors}
\label{sec4}
%
In this section we switch back to the original zigzag codes and consider the problem of correcting errors.  

First, we consider the  $(k+2,k)$ zigzag code. Since the code has minimum Hamming distance $3$ by the MDS property, if there is a node erasure and a node error, then one cannot expect to recover the information. However, we will see that zigzag code can correct if certain subset of the node elements are erroneous. As a result, zigzag code has the capability to correct element errors beyond the minimum Hamming distance. 
We will first study the scenario where an entire node erasure and a single element error happen at the same time, and remark on its generalization to correcting a subset of element errors.

Second, we consider the $(k+r,k)$ zigzag code. Since the the minimum Hamming distance is $r+1$, one can recover from $\lfloor \frac{r+1}{2} \rfloor \ge 1$ errors, for $r \ge 2$.
We will show a simplified error correcting algorithm for one node error, which is the most common case for node errors.

\subsection{Node Erasure and Element Error}
Consider a $(k+2,k)$ general zigzag code generated by \emph{distinct} binary vectors $T=\{v_0,v_1,\dots,v_{k-1}\}$. 
We assume that the erasure and the error are in different systematic columns, and there is at most a single element error in the systematic part of the array. The code has two parities and $p=2^m$ rows for some integer $m$, and the zigzag permutations are $f_{j}(x)=x+v_j, x \in [0,2^m-1], j\in [0,k-1]$. The original and the erroneous information array are denoted by $(a_{i,j})$,  $(\hat{a}_{i,j})$, respectively. 
We use $r_i,z_i$ to denote the $i$-th element in the 0th parity and 1st parity, respectively.
Let the coefficients of the 0th parity be all ones, and let the coefficient of the 1st parity corresponding to $a_{i,j}$ be $\beta_{i,j}$, which is non-zero. More specifically, by Construction \ref{cnstr5}, for all $i \in [0,2^m-1]$,
\begin{align*}
r_i &= \sum_{j=0}^{k-1} {a}_{i,j}, \\
z_i &= \sum_{j=0}^{k-1} \beta_{f_j^{-1}(i),j} {a}_{f_j^{-1}(i),j}.
\end{align*}

We will use the following fact.
By computing the determinant of every $2 \times 2$ sub-matrix of \eqref{eqG}, one can show that the zigzag code is MDS if and only if $\{beta_{i,j}\}$ are non-zero, and
\begin{equation}\label{eq4}
  \beta_{r,t}\beta_{r',t} \neq \beta_{r,j}\beta_{r',j}
\end{equation}
for all $r'=r+v_i+v_j$ (A detailed proof can also be found in \cite{Tamo2}).

{\bf Notation:} Let $x_0,x_1,\dots,x_{p-1} \in \mathbb{F}$. 
Denote $f(x_0,x_1,\dots,x_{p-1})=(x_{f(0)},x_{f(1)},\dots,x_{f(p-1)})$ for a permutation $f$ on $[0,p-1]$. 
For two vectors $B=(b_0,\dots,b_{p-1}),X=(x_0,\dots,x_{p-1}) \in \mathbb{F}^p$, denote by $B \circ X =(b_0x_0,\dots,b_{p-1}x_{p-1})$ the point-wise product.

\begin{algorithm} 
Suppose systematic column $t$ is erased, and there is at most one element error in the remaining systematic array. Denote by $B=(\beta_{0,t},\beta_{1,t},\dots,\beta_{p-1,t})$ the vector of zigzag coefficients corresponding to column $t$.
\begin{algorithmic}[1]
\For{$i \in [0,2^m-1]$}
	\State Compute the syndromes $S_{0}=(s_{0,0},s_{1,0},\dots,s_{2^m-1,0})$ and $S_{1}=(s_{0,1},s_{1,1},\dots,s_{2^m-1,1})$:
	\State $s_{i,0} \gets \sum_{j \neq t} \hat{a}_{i,j} - r_i,$
	\State $s_{i,1} \gets \sum_{j \neq t} \beta_{f_j^{-1}(i),j} \hat{a}_{f_j^{-1}(i),j} -z_i.$
\EndFor
\State $X  \gets  B \circ S_0.$ \label{l6}
\State $Y \gets f_t(S_1).$ \label{l7}
\State $W \gets X-Y.$ Denote $W=(w_0,\dots,w_{p-1}).$ 
\If{$W=0$}
	\State There is no element error. Assign column $t$ as $-S_0$. 
\Else
	\State Find two rows $r,r'$ such that $w_r, w_{r'}$ are nonzero. \label{l12}
	\State Find the unique $j$ such that $v_j=r+r'+v_t$. The error is in column $j$. \label{l13}
	\If {$\frac{w_r}{w_{r'}} = -\frac{\beta_{r,t}}{\beta_{r,j}}$} \label{l14}
		\State The error is at row $r$, and $a_{r,j}  \gets \hat{a}_{r,j}-\frac{w_{r}}{\beta_{r,t}}$. 
		\State Assign column $t$ as $-S_0$ for all elements except $a_{r,t}\gets -s_{r,0}+\frac{w_{r}}{\beta_{r,t}}$ 
	\Else
		\If{$\frac{w_r}{w_{r'}} = -\frac{\beta_{r',j}}{\beta_{r',t}}$}
		\State The error is at row $r'$, and $a_{r',j}  \gets \hat{a}_{r',j}-\frac{w_{r'}}{\beta_{r',t}}$. 
		\State Assign column $t$ as $-S_0$ for all elements except $a_{r',t}\gets -s_{r',0}+\frac{w_{r'}}{\beta_{r',t}}$ 
		\EndIf
	\EndIf \label{l18}
\EndIf
\end{algorithmic}
\caption{Algorithm to recover from a column erasure and an element error in the $(k+2,k)$ general zigzag code.}
\label{alg1}
\end{algorithm}

The procedure to recover 1 node erasure plus 1 element error is described in Algorithm \ref{alg1}.
We next show that its correctness.

\begin{thm}
Algorithm \ref{alg1} can correct a node erasure and a systematic element error for a $(k+2,k)$ MDS general zigzag code.
\end{thm}
\begin{IEEEproof}
Suppose column $t$ is erased and there is an error at column $j$ and row $r$. Let $\hat{a}_{r,j}=a_{r,j}+e$.  Define $r'=r+v_t+v_j$. 
Suppose $X=(x_0,\dots,x_{p-1})$, $Y=(y_0,\dots,y_{p-1})$ are the vectors in Lines \ref{l6}, \ref{l7}, respectively.
It is easy to see that $x_i = y_i = -\beta_{i,t}a_{i,t}$ except when $i=r,r'$. Thus the pair $r,r'$ found in Line \ref{l12} is unique. Since the set of binary vectors $\{v_0,v_1,\dots,v_{k-1}\}$ are distinct, we can identify column $j$ given $t,r,r'$ in Line \ref{l13}. Moreover,we have
$$x_r=-\beta_{r,t}a_{r,t}+\beta_{r,t}e,$$
$$y_{r}=-\beta_{r,t}a_{r,t},$$
$$x_{r'}=-\beta_{r',t}a_{r',t},$$
$$y_{r'}=-\beta_{r',t}a_{r',t}+\beta_{r,j}e.$$
Therefore, the difference between $X$ and $Y$ is
$$w_r = x_r-y_r = \beta_{r,t}e,$$
$$w_{r'}=x_{r'}-y_{r'}=-\beta_{r,j}e.$$
And we can see that no matter what $e$ is, we always have
$$\frac{w_r}{w_{r'}} = -\frac{\beta_{r,t}}{\beta_{r,j}}.$$
Similarly, if the error is at row $r'$ column $j$, we will get
$$\frac{w_r}{w_{r'}} = -\frac{\beta_{r',j}}{\beta_{r',t}}.$$
Since the code is MDS, we know that \eqref{eq4} holds. Therefore, we can distinguish between the two cases of an error in row $r$ and in row $r'$.
\end{IEEEproof}

\begin{xmpl}
Consider the code in Figure \ref{fig1-p} generated by $T=\{e_0=0,e_1,e_2\}$. Suppose all of Column $C_0$ is erased. And suppose there is an error in the 0-th element in Column $1$. Namely, the erroneous symbol we read is $\hat{a}_{0,1}=a_{0,1}+e$ for some error $e \neq 0 \in \mathbb{F}_3$, see Figure \ref{fig2}. We can simply compute the syndrome, locate this error, and recover the original array. In particular, since the erased column corresponds to the zero vector, and all the coefficients in column $0$ are ones, $X=S_0,Y=S_1$. For $i \in [0,3]$, we compute $W$, and get zeros in all places except row $0$ and $2$, which satisfy $0+2=(0,0)+(1,0)=(1,0)=e_1+e_0$. Therefore, we know the location of the error is in column $1$ and row $0$ or $2$. But since $w_0=w_2$, we know the error is in $\hat{a}_{0,1}$ (If $w_0=-w_2$, the error is in $\hat{a}_{2,1}$).
\end{xmpl}

\begin{figure}
  \centering
  \begin{tabular}{|c|c|c|c|c|c||c|c|c|}
  \hline
  ~& $C_0$ &	  $C_1$ & $C_2$ & $P_0$ & $P_1$ & $S_0$ & $S_1$ & $W=S_0-S_1$ \\
  \hline
  0 &  & $a_{0,1}+e$ &  $a_{0,2}$ & $r_0=a_{0,0}+a_{0,1}+a_{0,2}$ & $z_0=a_{0,0}+a_{2,1}+a_{1,2}$ 
   & $-a_{0,1}+e$ & $-a_{0,0}$ & $w_0=e$\\
  \hline
  1 &  & $a_{1,1}$ &  $a_{1,2}$ & $r_1=a_{1,0}+a_{1,1}+a_{1,2}$ & $z_1=a_{1,0}+a_{3,1}+2a_{0,2}$ 
  & $-a_{1,0}$   & $-a_{1,0}$ & $w_1=0$\\
  \hline
  2 &  & $a_{2,1}$ &  $a_{2,2}$ & $r_2=a_{2,0}+a_{2,1}+a_{2,2}$ & $z_2=a_{2,0}+2a_{0,1}+2a_{3,2}$ 
  & $-a_{2,0}$   & $-a_{2,0}+2e$ & $w_2=-2e=e$\\
  \hline
  3 &  & $a_{3,1}$ &  $a_{3,2}$ & $r_3=a_{3,0}+a_{3,1}+a_{3,2}$ & $z_3=a_{3,0}+2a_{1,1}+a_{2,2}$ 
  & $-a_{3,0}$   & $-a_{3,0}$ & $w_3=0$\\
  \hline
  \end{tabular}
  \caption{An erroneous array of the $(5,3)$ zigzag code. There is a node erasure in Column $C_0$ and an element error in Column $C_1$. $S_0,S_1$ are the syndromes.}\label{fig2}
\end{figure}



We make a few remarks about properties and extensions of our algorithm.
\begin{itemize}
\item
Consider the optimal zigzag code. In practice, when we are confident that there are no element errors besides the node erasure, we can use the optimal rebuilding algorithm and access only half of the array to rebuild the failed node. However, we can also try to rebuild this node by accessing the other half of the array. Thus we will have two recovered version for the same node. If they are equal to each other, there are no element errors; otherwise, there are element errors. Thus, we have the flexibility of achieving optimal rebuilding ratio or correcting extra errors.

\item
When node $t$ is erased and more than one element in column $j \neq t$ and rows $R \subseteq [0,q-1]$ are erroneous, following the same techniques as Algorithm \ref{alg1}, it is easy to see that the code is able to correct systematic errors if (i) $R \cup (R+v_j) \neq S \cup (S+v_{i}) $ for any set of rows $S \subseteq [0,q-1]$ and any column $i \notin \{ j,r\}$, and (ii) $r' \neq r+v_j+v_t$ for any $r,r' \in R$. For example, consider the optimal zigzag code in Theorem \ref{thm_cnstr1} with $m=3,k=4,r=2$. If node $t=0$ is erased, and elements in column $j=1$ and rows $R=\{0,7\}$ are erroneous, then our algorithm in Lines \ref{l12}, \ref{l13} will identify that there are errors in rows $\{0,4,3,7\}$, and only $e_j=e_1$ satisfies  $S \cup (S+v_j)=\{0,4,3,7\}$ for some $S$. Then using Lines \ref{l14} to \ref{l18} we can identify rows $R$ as erroneous and correct them.

\item
When the code has more than two parities, the zigzag code can again correct element errors exceeding the minimum Hamming distance. To detect errors, one can either compute the syndromes, or rebuild the erasures multiple times by accessing different $e/r$ parts of the array, where $e$ is the number of node erasures.

\item
Finally, it should be noted that if the node erasure or the single error happen in a parity column, then we can not correct them in the $(k+2,k)$ code.
\end{itemize}

\subsection{Node Error}
Next, we will discuss decoding algorithms of zigzag codes with $r$ parities in case of a column error.

Let $\cC$ be an $(k+r,k)$ general zigzag code defined by Construction \ref{cnstr5}. The code has information array size $2^m \times k$. Let the zigzag permutations be $f_j$, $j \in [0,k-1]$, which are not necessarily distinct.
Let the stored information be $a_0,a_1,\dots,a_{k-1}$ and parities be $b_0,b_1,\dots,b_{r-1}$, where each $a_i$ or $b_j$ corresponds to a node and is a column vector of length $r^m$. Let the erroneous nodes be $\hat{a}_0,\dots,\hat{a}_{k-1},\hat{b}_0,\dots,\hat{b}_{r-1}$.
Let the encoding matrix corresponding to systematic node $i$ and parity $l$ be $P_i^l$.

The procedure to correct a column error is shown in Algorithm \ref{alg2}.
We first compute the syndromes (Line \ref{p2}), then for a systematic node error we locate the error position using only syndromes from parity 0 and 1 (Line \ref{p5}), and at last correct the error (Line \ref{p6}).

Notice that the muliplication of the encoding matrix $P_j^l$ in Line \ref{p2} is only a permutation and muliplying coefficients. Moreover, the permutations $f_i$'s only change one bit of the row indeces, if we consider the optimal zigzag code in Theorem \ref{thm_cnstr1}. Therefore the algorithm can be easily implemented.

\begin{algorithm} 
\begin{algorithmic}[1]
\For {$l \in [0, r-1]$}
	 \State $S_l \gets \hat{b}_l - \sum_{i=0}^{k-1} P_i^l \hat{a}_i $ \label{p2}
\EndFor
\If {$S_l=0$ for all $l \in [0,r-1]$} 
	\State There is no error. 
\Else	\If{$S_l \neq 0$, for one $l$} 
			\State There is an error in parity $l$. $b_l \gets \hat{b}_l - S_l$. \label{p4}
		\Else
			\State Find the unique $j \in [0,k-1]$ such that $P_jS_0 = S_1$. \label{p5}
			\State $a_j \gets \hat{a}_j+S_0$. \label{p6} 
		\EndIf
\EndIf
\end{algorithmic}
\caption{Decode a node error in the $(k+r,k)$ general zigzag code.}
 \label{alg2} 
\end{algorithm}

If there is only one error, the above algorithm is guaranteed to find the error location and correct it, as the following theorem states.

\begin{thm}
Algorithm \ref{alg2} can correct one node error for an MDS general zigzag code.
\end{thm}
\begin{IEEEproof}
Suppose there is error in the parity node $l$, then clearly line \ref{p4} recovers it.
Suppose there is error in the systematic column $j$, and $\hat{a}_j = a_j + E$, for error vector $E$. Thus the first two syndromes are
$$S_0=-E,$$
$$S_1=-P_j E = P_j S_0.$$
Thus column $j$ will be found in Line \ref{p5}.
Next we  show that any other column will not be found in Line \ref{p5}. Namely, for any $t \neq j$, $P_tS_0 \neq S_1$.
Since the zigzag code is MDS, it can correct two erasures in nodes $j,t$. Therefore, the following matrix should be invertible:
$$\left(
\begin{matrix}
I & I \\
P_j & P_t
\end{matrix}
\right).
$$
Hence $P_j-P_t$ is invertible. Therefore, $-(P_j-P_t)E \neq 0$ since $E \neq 0$, namely, $S_1 \neq P_tS_0$.
\end{IEEEproof}

%
\section{Summary}
\label{sec6}
%
In this paper, we proved a tight information-theoretic lower bound on the rebuilding ratio $e/r$ for $e$ node failures and $r$ parities, and gave explicit rebuilding algorithms of optimal zigzag codes achieving this bound.
We also presented constructions of MDS array codes that achieve the optimal rebuilding ratio $1/r$ for an arbitrary node failure. The new codes are constructed using permutation matrices and improve the efficiency of the rebuilding access. Moreover, we considered the correction of errors, both for an element and for a node, which was not very well studied in erasure coding for distributed storage.

Now we mention a couple of open problems. 
First, if there are $k=m-1$ systematic nodes and $r$ parity nodes, then our code in Section \ref{sec5} has $p=r^m$ rows. Thus, the code dimension $k$ is quite small compared to the number of rows $p$, which limits the total number of nodes in the distributed storage network for a given storage size of every node. Given the number of rows $p=2^m$, it is theoretically interesting and practically important to study whether it is possible to find codes with a larger $k$. For example, when $r=2$, we know a construction with $r^m$ rows and $k=m$ systematic nodes:
$$A_j^0=\left(\begin{matrix}
  I & 0 \\
  p_j & I
\end{matrix}
\right),
A_j^1=\left(\begin{matrix}
  I & p_j \\
  0 & I
\end{matrix}
\right).
$$
Here $A_j^0,A_j^1$ are the encoding matrices for systematic node $j$ and parities $0,1$ respectively, and we can take all $j \in [1,m]$. This code has one more information column than Construction \ref{cnstr2}, and one can show that it achieves optimal rebuilding ratio as well. 


Besides, the zigzag code and the code in Section \ref{sec5} only specifies the finite field sizes when the number of parities is small, i.e., $r=2,3$. Hence it is useful to study the minimum required field sizes of these constructions, and to construct explicit codes using small finite field sizes.

Finally, the problem of correcting node erasures together with element errors in multiple nodes for codes with $r >2$ is important especially for applications in SSD-based storage systems.


\bibliographystyle{IEEEtranS}
\bibliography{mybib}

\end{document}